\documentclass[12pt]{article}
\usepackage{amsfonts,amsmath}
\usepackage{amssymb}

\newcommand{\dr}{{{\rm d}}}
\newcommand{\dz}{{{\rm d_z}}}
\newcommand{\hf}{\circ}
\renewcommand{\theequation}{\thesection.\arabic{equation}}

\makeatletter \@addtoreset{equation}{section} \makeatother

{\vspace{3mm} }

\def\al{\alpha}

\def\*{\star}
\def\e{\mathbf{e}}
\def\E2{\mathbf{E}}

\newcommand{\be}{\begin{equation}}
\newcommand{\ee}{\end{equation}}
\newcommand{\bee}{\begin{eqnarray}}
\newcommand{\beee}{\begin{array}}
\newcommand{\eee}{\end{eqnarray}}
\newcommand{\eeee}{\end{array}}
\newcommand{\gb}{\beta}
\newcommand{\gga}{\gamma}

\newcommand{\gd}{\delta}

\newcommand{\gk}{\varkappa}
\newcommand{\gep}{\epsilon}

\newcommand{\gs}{\sigma}
\newcommand{\go}{\omega}

\newcommand{\dal}{\dot \alpha}
\newcommand{\dgb}{\dot \beta}
\newcommand{\dgga}{\dot \gamma}

\newcommand{\nn}{\nonumber}

\newcommand{\p}{\partial}

\newcommand{\ff}{\frac}

\begin{document}
\begin{flushright}
FIAN/TD/12-2022\\
\end{flushright}

\vspace{0.5cm}
\begin{center}
{\large\bf On holomorphic sector of higher-spin theory}

\vspace{1 cm}

\textbf{V.E.~Didenko}\\

\vspace{1 cm}

\textbf{}\textbf{}\\
 \vspace{0.5cm}
 \textit{I.E. Tamm Department of Theoretical Physics,
Lebedev Physical Institute,}\\
 \textit{ Leninsky prospect 53, 119991, Moscow, Russia }\\

\par\end{center}

\begin{center}
\vspace{0.6cm}
e-mail: didenko@lpi.ru \\
\par\end{center}

\vspace{0.4cm}

\begin{abstract}
\noindent Recent investigation of the locality problem for
higher-spin fields led to a vertex reconstruction procedure that
involved elements of contraction of the original Vasiliev
interaction algebra. Inspired by these results we propose the
Vasiliev-like generating equations for the holomorphic higher-spin
interactions in four dimensions based on the observed contracted
algebra. We specify the functional class that admits evolution on
the proposed equations and brings in a systematic procedure of
extracting all-order holomorphic vertices. A simple consequence of
the proposed equations is the space-time locality of the gauge
field sector. We also show that vertices come with a remarkable
shift symmetry.

\end{abstract}
\newpage
\tableofcontents
\newpage
\section{Introduction}
\subsection{Higher-spin locality problem}

A notoriously difficult open problem is the structure of
(non)-local higher-spin (HS) gauge interactions
\cite{Vasiliev:1999ba}-\cite{Ponomarev:2022vjb}. The question is
whether the theory is local and if not what a missing locality
should be replaced with? A certain kind of non-locality is
expected due to a HS symmetry that mixes fields of arbitrary large
spins via higher space-time derivatives \cite{Bengtsson:1983pd},
\cite{Berends:1984rq}, \cite{FrVas}. This results in that while a
free theory carries no more than two derivatives, the cubic
interaction $V_{s_1, s_2, s_3}$ involves higher order ones. Their
number grows with spin being bounded from above by
$\#\p=s_1+s_2+s_3$, \cite{Metsaev:2005ar}. Since the HS spectrum
is unbounded in $d>3$ this implies the theory is non-local beyond
free level. A cubic example illustrates however that this type of
non-locality is under full control once interaction is restricted
to fixed three spins. The corresponding vertex contains finitely
many derivatives and therefore is local. Such vertices can be
treated using standard field theory tools. In particular, they can
be recovered  from a free CFT \cite{Sleight:2016dba} by inverting
the Witten diagrams of $HS/O(N)$ duality \cite{Klebanov:2002ja},
\cite{Sezgin:2003pt}, \cite{Leigh:2003gk}, \cite{Giombi:2011kc}
testing the Klebanov-Polyakov conjecture \cite{Klebanov:2002ja} at
this order along the way.

A natural question is whether a quartic HS vertex $V_{s_1, s_2,
s_3, s_4}$ is local or not. The holographic reconstruction of
scalar self-interaction $0-0-0-0$ from the $O(N)$ four-point
correlation function carried out in \cite{Bekaert:2015tva} has
sowed doubts \cite{Sleight:2017pcz} on the locality at this order
as the final result appeared to involve infinitely many space-time
derivatives even for fixed spins $s_i=0$\footnote{The holographic
reconstruction of \cite{Bekaert:2015tva} should be considered more
like a signal of a potential problem rather than a derivation of
the vertex. Indeed, the appearance of infinitely many derivatives
affects the Lagrangian analysis and calls for a precise definition
of its functional class. For example, by allowing $\ff1\Box$ in
the Lagrangian one no longer can set apart bulk and boundary
integral contributions.}. Although the non-local result of
\cite{Bekaert:2015tva} can not be treated as a solid prove of
irremovable non-locality, it raises a big concern on the existence
of a HS theory in the form of a local field theory. Apart from the
assumption of the holographic duality at the $4pt$ -- level
itself, a weak spot of \cite{Sleight:2017pcz} from the CFT side is
an infinite series of single trace conformal blocks which
singularity is not fully understood (see \cite{Ponomarev:2017qab}
for analysis of this issue). On the $AdS$ side the
non-commutativity of the $AdS$ derivatives $[D,D]\sim R^{-2}$ may
affect the reconstruction procedure as well. It should be also
stressed that the inherent field redefinition ambiguity that may
change a local form of the vertex into a non-local one is very
hard to take into account within the holographic reconstruction.
This is why it is important to investigate the locality issue
using the $AdS/CFT$ independent tools.

\subsection{Unfolded approach}

The language of differential forms that take values in the HS
algebra is at the core of the unfolding formalism for higher spins
\cite{Vasiliev:1988sa}. Its classical differential equations of
motion\footnote{See \cite{Misuna:2022cma} for an application of
the unfolded dynamics at quantum level.} are of first order, while
gauge invariance is inbuilt through a formal consistency. The
price one pays for an apparent simplicity is a necessity of an
infinite number of auxiliary fields. For the purely symmetric
bosonic fields in $d$ dimensions the spectrum of these fields is
governed by the Eastwood-Vasiliev HS algebra
\cite{Eastwood:2002su}, \cite{Vasiliev:2003ev}. These are the
two-row traceless $o(d-1,2)$ Young diagrams. However, despite the
full nonlinear equations of motion for these gauge fields are
known in a closed form \cite{Vasiliev:2003ev}, their analysis is
not yet well elaborated beyond free level.

While the HS locality problem can be addressed in any space-time
dimension, a particular instance of $d=4$ has a great advantage as
compared to general $d$. It is in this case that the HS algebra
admits a very simple realization in terms of the two-component
spinors \cite{Vasiliev:1999ba} reducing it to the enveloping of
$Y_{A}=(y_{\al}, \bar y_{\dal})$, $\al, \dal=1,2$ modded by
\be\label{HS}
[y_{\al}, y_{\gb}]_\star=2i\gep_{\al\gb}\,,\qquad [y_{\al}, \bar
y_{\dgb}]_\star=0\,,\qquad [\bar y_{\dal}, \bar
y_{\dgb}]_\star=2i\gep_{\dal\dgb}\,,
\ee
where star product $\star$ can be chosen to be the Moyal one
\be\label{exp}
f\star g=f(y, \bar y)
e^{i\gep^{\al\gb}\overleftarrow{\p}_\al\overrightarrow{\p}_\gb+
i\gep^{\dal\dgb}\overleftarrow{\p}_{\dal}\overrightarrow{\p}_{\dgb}}
g(y, \bar y)
\ee
and $\gep_{\al\gb}=-\gep_{\gb\al}$ is the $sp(2)$ invariant form
(same for $\gep_{\dal\dgb}$). Note that the star product is given
by a tensor product of two pieces -- the holomorphic that acts on
$y$ and the anti-holomorphic one acting on $\bar y$
\be\label{barexp}
f\bar\star\, g=f(y, \bar y) e^{
i\gep^{\dal\dgb}\overleftarrow{\p}_{\dal}\overrightarrow{\p}_{\dgb}}
g(y, \bar y)\,.
\ee
The $sl(2,\mathbb{C})$ dictionary implies that the two-row
$o(3,2)=sp(4,\mathbb{R})$ Young diagrams are mere symmetric
multispinors in dotted and undotted indices and therefore can be
packed into generating functions as formal polynomials of $y$ and
$\bar y$. This way one introduces a space-time one-form $\go(y,
\bar y)$ and a zero-form $C(y, \bar y)$. The unfolded HS equations
are\footnote{The commutator on the l.h.s of \eqref{eq2} is in the
twisted-adjoint representation of HS algebra rather than in the
adjoint as \eqref{eq2} suggests. This is usually achieved by
introducing outer Klein operators $k$ and $\bar k$ within the
field dependence \cite{Vasiliev:1999ba} that implement the twisted
automorphism. We omit those for brevity.}
\begin{align}
&\dr_x\go+\go\star\go=\Upsilon(\go, \go, C)+\Upsilon(\go, \go, C, C)+\dots\label{eq1}\\
&\dr_x C+[\go, C]_\star=\Upsilon(\go, C, C)+\Upsilon(\go, C, C,
C)+\dots\label{eq2}
\end{align}
Let us specify some important properties of these equations
\begin{itemize}
\item The space-time derivative appears in the form of de-Rahm
differential in \eqref{eq1}-\eqref{eq2}. Particularly, no manifest
derivatives are there in $\Upsilon$'s. It does not of course imply
that the space-time vertices contain no derivatives, rather it
says that the first derivative of certain fields is expressed in
terms of other fields on-shell. This is a manifestation of the
fact that $\go(y,\bar y)$ and $C(y,\bar y)$ along with the
physical fields contain plenty of the auxiliary ones that get
expressed via the former on \eqref{eq1}-\eqref{eq2}. Therefore,
any vertex $\Upsilon$ accumulates space-time derivatives when
expressed in terms of physical fields.

\item Vertices $\Upsilon$ can be found order by order in $C$ by
inspecting the integrability requirement $\dr_x^2=0$. The
resulting relations stem from the HS algebra action present on the
l.h.s of \eqref{eq1}-\eqref{eq2}. This procedure is naturally
defined up to a field redefinition $\go\to\go+f_{\go}(\go,
C,\dots, C)$ and $C\to C+f_C(C,\dots, C)$. This ambiguity is at
the core of the locality problem. A systematic way of extracting
$\Upsilon$'s is given by the all order Vasiliev equations
\cite{Vasiliev:1992av}. Their remarkable feature is that being
field redefinition independent they allow for an all order
analysis of a functional class corresponding to a one or another
field frame. This makes the Vasiliev approach very suitable for
the locality analysis.

\item The vacuum equations correspond to $\Upsilon=0$ which are
satisfied by a nontrivial $AdS_4$ background connection $\go_0$ or
any other HS flat connection. The linearized equations correspond
to the vanishing of all $\Upsilon(\go, C,\dots, C)=0$ and a
nontrivial $\Upsilon(\go, \go, C)$. We note that the linearized
vertex contains some information on the cubic interactions
(quartic in Lagrangian counting). This happens because the HS
algebra action on the l.h.s of \eqref{eq1}-\eqref{eq2} already
stores some details of quadratic interaction. This makes the
unfolded deformation procedure crucially different from the
standard Noether one which can be trivialized by non-local field
redefinitions. In the unfolded case there are no field
redefinitions that trivialize cubic coupling on a general HS
vacuum because this coupling partly comes with the free one.

\item Vertices of \eqref{eq1}-\eqref{eq2} depend on a free
constant complex parameter $\eta$, which unless $\eta=1$ or
$\eta=i$ breaks parity of HS interactions. From the boundary side
$\eta$ arguably interpolates between free (critical) boson and
free (critical) fermion via the $3d$ Chern-Simons interaction
manifesting the so-called $3d$ bosonization \cite{Giombi:2011kc},
\cite{Aharony:2012nh}. $\Upsilon$'s depend on $\eta$ in the
following fashion
\begin{align}
&\Upsilon(\go, \go, \underbrace{C,\dots, C}_{N})=\sum_{k=0}^{N}
\eta^k\bar\eta^{N-k}\Upsilon_{k}(\go, \go, \underbrace{C,\dots,
C}_{N})\,,\label{holW}\\
&\Upsilon(\go, \underbrace{C,\dots, C}_{N})=\sum_{k=0}^{N-1}
\eta^k\bar\eta^{N-k-1}\Upsilon_{k}(\go, \underbrace{C,\dots,
C}_{N})\,,\label{holC}
\end{align}
where $\Upsilon_k$ are $\eta$ -- independent.

\item Since $\eta$ and $\bar\eta$ appear as formally independent
variables in any of consistency relations for
\eqref{eq1}-\eqref{eq2} one can set, say,
$\bar\eta=0$\footnote{Such a reduction ruins the reality condition
implying that the (anti)holomorphic sector is essentially
complex.}. The resulting system remains consistent. This way one
reduces HS equations down to the holomorphic sector, originally
called the self-dual \cite{Vasiliev:1992av} and sometimes chiral
\cite{Iazeolla:2007wt}. As it follows from the early analysis by
Metsaev \cite{Metsaev:1991mt}, \cite{Metsaev: PhD} based on the
light-cone approach  (see e.g., \cite{Ponomarev:2017nrr} for a
recent account) this sector is fixed by a cubic approximation
receiving no higher order corrections on the Minkowski background.
One does not expect a similar behavior within the covariant
approach on the $AdS$ or general HS background though.
\end{itemize}

\subsection{Unfolded view of (non)-locality}
As it was stressed, the unfolded approach somehow obscures the
derivative structure of interaction in terms of the auxiliary
fields. To get at the derivative map let us have a closer look at
the generating fields $\go$ and $C$ from \eqref{eq1}-\eqref{eq2}.
For a detailed analysis we refer the reader to
\cite{Gelfond:2019tac}, \cite{Vasiliev:2022med}. A spin $s$ field
is singled out by the following conditions
\begin{align}
&\left(y^{\al}\ff{\p}{\p y^{\al}}+\bar y^{\dal}\ff{\p}{\p \bar
y^{\dal}}\right)\go=2(s-1)\,\go\,,\label{w}\\
&\left(y^{\al}\ff{\p}{\p y^{\al}}-\bar y^{\dal}\ff{\p}{\p \bar
y^{\dal}}\right)C=\pm2s\, C\label{C}\,,
\end{align}
where $\go$ contains gauge degrees of freedom, while $C$ accounts
for the gauge invariant combinations (spin $s$ analogs of
Maxwell's tensor). Note, that as follows from \eqref{w}, fields
$\go_{\al_{1}\dots\al_{m}, \dgb_{1}\dots\dgb_{n}}$ describe spin
$s$ iff $m+n=2(s-1)$, which says that both $m$ and $n$ are bounded
by the value of spin. This gives one a spin increasing but a
finite set of fields for a given gauge field. In particular, the
Fronsdal spin $s$ component is stored in
$\go_{\al_1\dots\al_{(s-1)}, \dgb_1\dots\dgb_{(s-1)}}$, while the
rest fields are auxiliary. The approximate derivative map that
relates these to the Fronsdal ones at free level is as follows
\be\label{waux}
\go_{[m],[n]}\sim \left(\ff{\p}{\p
x}\right)^{|m-s+1|}\go_{[s-1],[s-1] }\,,
\ee
where by $[m]$ we denote a group of $m$ indices and the spinor
version of $x_{a}\sim x_{\al\dal}$. Things are different with $C$
as from \eqref{C} it follows that a spin $s$ component
$C_{\al_{1}\dots\al_{m}, \dgb_{1}\dots\dgb_{n}}$ satisfies
$|m-n|=2s$ and therefore there are infinitely many components for
a given spin. The physical one is purely (anti)-holomorhic
$C_{\al_{1}\dots\al_{2s}}$ and $C_{\dgb_{1}\dots\dgb_{2s}}$.
Again, one can estimate now how the auxiliary components are
expressed via the physical ones
\be\label{Caux}
C_{[m],[n]}\sim \left(\ff{\p}{\p x}\right)^{\min(m,n)}C_{[2s]}\,.
\ee
\subsubsection{Spin locality and ultra-locality}
It is clear now where a potential non-local obstruction may come
from. Whenever one encounters infinitely many pairs of contracted
dotted and undotted indices in $C$'s (mind to keep spins fixed by
\eqref{C}) the corresponding contribution is non-local by argument
\eqref{Caux}. An example of such a non-locality is easy to devise
\be
\sum_{n} a_n C_{\al_1\dots\al_{2s_1}\gga_1\dots\gga_n,
\dgga_1\dots\dgga_n}C_{\gb_1\dots\gb_{2s_2}}{}^{\gga_1\dots\gga_n,
\dgga_1\dots\dgga_n}\,,
\ee
provided coefficients $a_n$ above have infinitely many nonzero
values. Here we have a non-local interactions of spin $s_1$ and
spin $s_2$.

One can define the notion of {\it spin locality}
\cite{Gelfond:2018vmi}, \cite{Gelfond:2019tac} for the vertices
from \eqref{eq1}-\eqref{eq2}. We call $\Upsilon$ a spin local if
being restricted to fixed spins it contains no more than a finite
amount of contractions between different $C$'s. By a single
contraction we assume a contraction of a one pair of dotted and
undotted indices, e.g. $f_{\al\dgb}g^{\al\dgb}$.

Few comments are now in order. While one may expect that the
notion of spin locality is equivalent to the space-time locality,
this may not be necessarily the case. As is carefully analyzed in
\cite{Gelfond:2019tac} the two notions are equivalent if the
number of physical fields is finite. The equivalence was not shown
otherwise (see however \cite{Vasiliev:2022med} where this problem
has been recently detailed). Another comment is that there is no
need to bother of contractions between two $\go$'s or between
$\go$ and $C$ since once restricted to a given spin, $\go_{s}$
becomes polynomial thanks to \eqref{w} and therefore such
contractions are always spin local.

A very important concept is the {\it spin ultra-locality}
\cite{Didenko:2018fgx} which is known to persist at lower orders
in $\Upsilon(\go, \go, \bullet)$ from \eqref{eq1}.
$\Upsilon(\go^{t_1}, \go^{t_2}, \bullet)$ is called a spin
ultra-local if the number of different index contractions is
bounded for all spins in its $C\dots C$ -- part (for fixed spins
$t_{1,2}$). In other words, the vertex dependence on field
$C^{s_i}$ must be organized via its physical (primary) component
or the auxiliary (descendant) ones which depth can not grow with
spin $s_i$. In practice this implies that the structure of such a
vertex in its $C$ dependence is roughly the following
\be
\Upsilon(\go^{t_1}, \go^{t_2}, C^{s_1},\dots, C^{s_n})\sim
\Phi(y,\p_{y_i}; \bar y, \p_{\bar y})\,C^{s_1}(y_1, \bar y)\dots
C^{s_n}(y_n, \bar y)\Big|_{y_i=0}\,,
\ee
where $\Phi$ depends polynomially on $y$ and $\p_{y_i}$ such that
its degree is independent of $s_i$. Note that the dependence on
$y$ (or similarly $\bar y$) drops off from $C$'s in the
ultra-local case. So, while a spin-local expression can not
contain infinitely many contractions for fixed spins, in the spin
ultra-local case the number of such contractions in addition is
bounded for any spins that enter $C$'s. An example of an
ultra-local 'vertex' comes already from the free equations, where
it has the form of the so called central on-mass-shell theorem
\cite{Vasiliev:1988sa}, \cite{Bychkov:2021zvd}
\be\label{omg}
\Upsilon(\go_0, \go_0, C)\sim \eta \bar
H^{\dal\dgb}\ff{\p^2}{\p\bar y^{\dal}\p\bar y^{\dgb}}C(0, \bar y)+
\bar\eta  H^{\al\gb}\ff{\p^2}{\p y^{\al}\p y^{\gb}}C(y, 0)\,,
\ee
where $H$ is the background $AdS$ two-form.

Let us also note that  the meaning of ultra-locality within the
Fronsdal formulation is not that transparent as the corresponding
ultra-local expressions when expressed in terms of physical fields
are just local and contain spin dependent derivatives. It is the
rate of an $s$--dependence that allows one distinguishing local
expressions from ultra-local ones.

\subsubsection{Star product and (non)-locality}
The global HS symmetry naturally introduces a certain non-locality
by mixing an infinite tower of gauge fields. Therefore, it is not
surprising that star product \eqref{exp} can be a source of
non-local terms. For example,
\be
C^{s_1}\star C^{s_2}=C^{s_1}(y, \bar
y)\,e^{i\overleftarrow{\p}\cdot \overrightarrow{\p}+
i\overleftarrow{\bar\p}\cdot\overrightarrow{\bar\p}} C^{s_2}(y,
\bar y)
\ee
is spin non-local for projection on any $s$ and for any fixed
$s_1$ and $s_2$, because it inevitably contains infinitely many
contractions of two types of indices. If however one leaves only
half of the star product nontrivial, say the one that acts on
$\bar y$, then a similar expression
\be\label{barstar}
C^{s_1}\bar\star\, C^{s_2}=C^{s_1}(y, \bar y)\,e^{
i\overleftarrow{\bar\p}\cdot\overrightarrow{\bar\p}} C^{s_2}(y,
\bar y)
\ee
is perfectly spin-local for it contains infinitely many
contractions of one type (dotted) only. We then note that the fate
of locality heavily relies on the structure of the underlying HS
algebra. If we are to keep the first relation in \eqref{HS}
nontrivial only by setting $[\bar y_{\dal}, \bar y_{\dgb}]=0$, the
resulting HS algebra delivers no apparent
non-localities\footnote{The space-time algebra made of billinears
in $y$'s is no longer the $AdS$ one in this case. Note, that it
does not reduce to the Minkowski either.}.

Let us look at the effect that star product \eqref{exp} produces
on (non)-locality of \eqref{eq1}-\eqref{eq2}. The left hand side
has the quadratic vertices governed by the HS symmetry. These are
$\go\star \go$ and $[\go, C]_\star$. Both are spin-local since $C$
appears once at best. The vertices on the right come from a HS
symmetry deformation, which emerges from the consistency
constraint $\dr^2_x=0$. For example, one easily extracts the
following condition\footnote{It is convenient to assume extra
Chan-Paton color indices on $\go$ and $C$. It allows one
considering different orderings of $\go$'s and $C$'s separately.}
at leading order by applying $\dr_x$ to \eqref{eq1} and
\eqref{eq2}
\be\label{ex}
\Upsilon_1(\go, \go, C)\star C=\Upsilon_0(\go, \go\star C,
C)-\Upsilon_0(\go\star \go, C, C)+\go\star\Upsilon_0(\go, C, C)\,,
\ee
where $\Upsilon_1$ and $\Upsilon_0$ come from 1-form \eqref{eq1}
and 0-form \eqref{eq2} sectors with all $\go$'s at mostly left
position in $\Upsilon_{0,1}$ . From \eqref{ex} it is clear that
$\Upsilon_0(\go, C, C)$ can not be local if the left hand side is
non-local. The latter having a star product of two $C$'s is
non-local unless $\Upsilon_1(\go, \go, C)$ brings no dependence on
$y$ or $\bar y$ in the argument of $C$. This happens if
$\Upsilon_1(\go, \go, C)$ is ultra-local. From this simple
analysis one concludes that the star product places very stringent
constraints for the vertices to be potentially local.
Particularly, some vertices have to be ultra-local to support
locality. An important lesson here is that even local field
redefinitions at a given leading order may result in non-local
vertices at the next-to-leading one. Indeed, $\Upsilon(\go, \go,
C)$ having only one $C$ is always local. It is its ultra-locality
however that gives $\Upsilon(\go, C, C)$ a local chance. The
nature of this phenomenon is directly related to an infinite
spectrum of HS fields and has no analog in the case of a finite
spectrum. In particular, this makes the canonical form of
\eqref{omg} crucially important for the cubic HS interactions to
be local.

\subsection{Present state of HS (non)-locality}
There are several approaches devoted to the HS interaction problem
in the literature (see e.g. \cite{Ponomarev:2022vjb} for a recent
pedagogical review and references therein). Some, like in
\cite{deMelloKoch:2018ivk}, \cite{Aharony:2020omh} give up on
locality from the very beginning assuming that once it breaks down
sooner or later in perturbations there is no need in hanging it
on. This point of view is supported in some sense by the
holographic quartic analysis \cite{Bekaert:2015tva} that unlike
the cubic one \cite{Sleight:2016dba} points out at non-locality.
The holographic approach pronounces HS theory non-local beyond
cubic order. It remains not clear so far how to define the theory
in the bulk without involving a conjectural holographic dual one
on this way. The matching of bulk and boundary pictures works fine
while local though \cite{Giombi:2009wh}, \cite{Misuna:2017bjb},
\cite{Sezgin:2017jgm}, \cite{Didenko:2017lsn} (see also
\cite{Lysov:2022nsv} for a very recent analysis at the cubic
order). At any rate, one has to accept that the holographic HS
reconstruction is paused until the quartic vertex is settled up.
An attempt to qualify the holographic non-locality arising at this
order is given in \cite{Ponomarev:2017qab}. Another feasible
option proposed in \cite{Vasiliev:2012vf}, \cite{Vasiliev:2022med}
is the proper boundary dual for the HS theory might be the
conformal HS theory rather than a vector model. If that is the
case then the holographic reconstruction should be revisited.

The approach based on the Vasiliev equations on the other hand
gives access to all order vertices in their unfolded form
\eqref{eq1}-\eqref{eq2}. What it does not tell is which field
frame one should pick for the interactions to be local or properly
non-local if the former is not possible. Being independent of any
particular field frame, the Vasiliev equations allow for all order
control of the functional class and therefore gives a tool for a
systematic analysis of (non)-locality. In a series of papers
\cite{Vasiliev:2016xui}-\cite{Gelfond:2021two},
\cite{Gelfond:2019tac}-\cite{Didenko:2018fgx}, the locality
problem was analyzed at first few interaction orders and some all
order statements were obtained. The results of these papers can be
summarized as follows.

Vertices that appear in \eqref{eq1} are shown to be local and
explicitly found\footnote{Note that the result of
\cite{Vasiliev:2016xui} invalidates the argument of
\cite{Skvortsov:2015lja} which states that if the Vasiliev
quadratic vertex is chosen to be non-local via the standard
homotopy resolution, there is no regular improvement that can
bring it into a local form.} at least up to $\Upsilon(\go, \go, C,
C)$, while those from \eqref{eq2} are local at least up to
$\Upsilon(\go, C, C, C)$ in the holomorphic sector ($\bar\eta=0$,
see \eqref{holC}). An explicit form of \eqref{eq2} is known for
$\Upsilon(\go, C, C)$, \cite{Didenko:2019xzz} and for a fragment
of $\Upsilon(\go, C, C, C)$, \cite{Gelfond:2021two}. Note that the
locality is proven to quite a high order with some of the vertices
from quintic interactions in the Lagrange counting. Still, neither
of those cover the complete quartic vertex as the fate of HS
quartic interaction remains indefinite. Particularly, the
$\eta\bar\eta$ -- part of $\Upsilon(\go, C, C, C)$ that completes
quartic $V_{0, s_1, s_2, s_3}$ is not yet analyzed.

In obtaining these results a number of important observations and
conjectures have been made. Let us specify those playing an
important role in our investigation.

\paragraph{Functional class} In the sequel we deal with the
Vasiliev generating equations for \eqref{eq1}-\eqref{eq2}. The
master fields entering these equations belong to a certain space
which is large enough to encompass both the local and (non)-local
field frames. Therefore, one has to specify it exactly in order to
provide a local frame if exists. While {\it a priori} it is not
clear how to do it, whatever this class is it should be closed on
the operations of the generating equations. This problem was first
analyzed in \cite{Vasiliev:2015wma}, where a certain  class was
proposed. The would be local class if exists should be a subclass
of this one as it was specified further in \cite{Gelfond:2019tac}.

\paragraph{Pfaffian locality theorem} The degree of non-locality
can be estimated by the important theorem \cite{Gelfond:2018vmi}
which states, that a perturbation theory prescribed by the
specific resolution operators reduces generic non-locality as
measured by rank of a certain Pfaffian. Particularly, this result
alone is sufficient to prove locality of vertex $\Upsilon(\go, C,
C)$. Although the result of \cite{Gelfond:2018vmi} is a crude
estimate of the exponential behavior of the non-local contractions
and does not answer whether the theory is local or not, it sets
the stage for a deeper analysis into HS interactions that meet the
Pfaffian locality theorem conditions.

\paragraph{Star-product re-ordering limit}In constructing the
resolutions that respect the aforementioned functional class, one
arrives at the homotopy operators that can be reinterpreted in
terms of conventional homotopies for the star-product re-ordered
Vasiliev equations followed by a certain limit of the re-ordering
parameter\footnote{The fact of equivalence upon re-ordering of the
standard homotopy with the properly shifted one was first noted
for $\gb=1$ in \cite{DeFilippi:2019jqq}.} $\gb\to-\infty$,
\cite{Didenko:2019xzz}. Specifically, the limit in question
represents a contraction of the large star-product algebra that
changes some commutation relations. This is a very important
result that shows that the (non)-locality is much affected by the
type of the large algebra. In different context a recent
investigation on the ordering (in)dependence can be found in
\cite{DeFilippi:2021xon}.

\paragraph{Holomorphic sector} The analysis of the HS vertex structure
in the holomorphic sector can be carried out independently from
the rest. Moreover, its status from the locality perspective is
crucial for the locality of the whole theory. By now it is known
that the holomorphic leading $\Upsilon(\go, \go, C)$ and
next-to-leading $\Upsilon(\go, \go, C, C)$ are ultra-local,
\cite{Didenko:2019xzz}. In \cite{Gelfond:2018vmi} an important
conjecture called the $z$ -- dominance lemma was proposed, which
if true allows to state a vertex locality without its manifest
calculation. The conjecture was proven at $CC$ -- order and
verified by a brute force calculation of $\Upsilon(\go, C, C, C)$
in \cite{Gelfond:2021two}. This technical condition if satisfied
serves as a useful guiding principle in probing locality. In
\cite{DK} we specify conditions for the $z$ -- dominance lemma and
prove its validity.  The all order spin locality conjecture of the
holomorphic sector has been put forward in \cite{Gelfond:2018vmi}.

\subsection{Structure of the paper}

The paper is organized as follows. In section \ref{goals} we
formulate our goals and summarize the main results. Section
\ref{vas} is devoted to the Vasiliev formulation of the HS
generating equations. In \ref{brief} we provide a brief reminder
of the original ideas, in section \ref{reord} we modify the
Vasiliev equations by introducing the $\gb$ -- ordering freedom,
in section \ref{lim} their $\gb\to-\infty$ contraction is
investigated. In section \ref{classes} we specify the proper
functional class and in section \ref{sec} analyze star products
within that class. Section \ref{gen} contains generating equations
for the HS holomorphic interactions. Its perturbation theory is
given in section \ref{pert} with the lower order examples of
interaction vertices in section \ref{vert}. The locality is proven
in section \ref{loc}, while an observation of a certain shift
symmetry is given in \ref{shift}. We conclude in section
\ref{concl}. The paper is supplemented with two appendices. In
Appendix A we derive our functional class. Appendix B provides
with a proof of the central projector identity \eqref{awe}.

\section{Goals and main results}\label{goals}

Our primary interest is the holomorphic sector. One would like to
understand whether the ultra-locality extends beyond
$\Upsilon(\go, \go, C, C)$, as well as whether the spin locality
extends beyond $\Upsilon(\go, C, C, C)$? In attacking this problem
we specify the functional class for the Vasiliev master fields.
The proper functions are chosen to be invariant under the
aforementioned $\gb$ -- re-ordering. This is a crucial observation
based on our experience of the earlier analysis of a few
interaction orders \cite{Didenko:2019xzz}. Such defined class
appears to be a subclass of the one proposed in
\cite{Gelfond:2019tac}, but unlike the latter it is not respected
by the original Vasiliev star product. Nevertheless it turns out
to be respected by a star product that emerges in the re-ordering
limit (contraction).

So, our goal is to come up with the Vasiliev-like generating
equations for the specified functional class. Should that be
possible one no longer needs quite a complicated homotopy
resolutions from \cite{Didenko:2019xzz}, which are designed to
bring fields to the proper class after star-product action.
Indeed, as shown in \cite{Didenko:2019xzz} the complicated
homotopy operators reduce to the standard contracting homotopy
upon the re-ordering followed by a contraction.

To reach this goal we scrutinize the $\gb$ -- ordering freedom of
the Vasiliev generating system and investigate the locality limit
$\gb\to-\infty$ at the level of equations of motion. Even though
the naive limit does exist, the resulting equations appear to make
no sense beyond lower orders as we show. At higher orders the new
star product brings infinities. This result is exceptionally
interesting for on the one hand we know that some higher order
vertices were indeed effectively calculated \cite{Didenko:2019xzz}
using the limiting star-product, but on the other, this limit
could not be taken at the level of the Vasiliev equations
directly. That suggests the existence of the different Vasiliev
type equations based on a new algebra.

We show that such equations do exist and indeed differ from those
obtained via a naive limit $\gb\to-\infty$. Equations
\eqref{dxWeq}-\eqref{Ceq} is the central result of our work. In
particular they do not contain a zero-form module $B$ which is
usually responsible for vertices $\Upsilon(\go,C,\dots, C)$. In
our case these missing vertices come up automatically in a way to
complete consistency. In addition to the standard Vasiliev case,
the mechanism that makes the whole system consistent rests on the
existence of a unique element called $\Lambda_0$ in the space of
$\dr z$ -- one-forms. Along with the usual Klein two-form the two
are the main building blocks of the generating system. We observe
a remarkable projector identity \eqref{awe} that involves
$\Lambda_0$. The identity is responsible for the consistency of
the whole system and is a major observation of the present paper.
The appearance of $\Lambda_0$ is something that makes difference
between our system and the standard Vasiliev equations.

A simple consequence of the obtained generating equations is the
all order ultra-locality of $\Upsilon(\go,\go, C,\dots, C)$ and
spin locality of $\Upsilon(\go,C,\dots, C)$ as we also show. Thus
we give a proof of the locality conjecture of
\cite{Gelfond:2018vmi}. Another important result is a shift
symmetry of the holomorphic vertices that holds to all orders. An
investigation of this symmetry was inspired by the structure lemma
of \cite{Gelfond:2018vmi} that underlies the Pfaffian locality
theorem and prescribes certain homotopy shifts in perturbation
theory. We show that precisely these shifts generate a symmetry of
HS vertices. This observation indicates a relation of the observed
symmetry with the HS locality.

In connection with our work let us point out a recent paper
\cite{Sharapov:2022awp} where the all order locality conjecture of
the holomorphic sector was also put forward. While the result of
\cite{Sharapov:2022awp} is as well based on the limiting star
product of \cite{Didenko:2019xzz}, unlike our approach the authors
engaged an accessory assumption called the 'duality map' that they
checked at lower orders. The duality map is supposed to help
reaching out $\Upsilon(\go,C,\dots, C)$ from $\Upsilon(\go,\go,
C,\dots, C)$. Lack of manifest consistency however forces the
authors examining the resulting vertices against consistency. This
way \cite{Sharapov:2022awp} explicitly checks few vertices up to
$\Upsilon(\go,\go, C, C)$ reproducing some of the earlier results
\cite{Didenko:2018fgx}, \cite{Didenko:2019xzz} and conjectures the
higher order ones. Our approach is free from any outside
assumptions on vertices and rests on the formal all order
consistency.

\section{Vasiliev equations}\label{vas}

\subsection{A brief reminder}\label{brief}
Here we would like to recall the basic elements behind the
Vasiliev equations. We are being somewhat sketchy there. Our
exposition is slightly different from the original
\cite{Vasiliev:1992av} (for reviews see \cite{Vasiliev:1999ba},
\cite{Bekaert:2004qos}, \cite{Didenko:2014dwa}) for a reason that
will be clear soon.

The idea of the generating system for the unfolded HS equations
\eqref{eq1}-\eqref{eq2} is to extend the dependence of fields
$\go$ and $C$ onto a larger space that includes variables
$Z_{A}=(z_{\al}, \bar z_{\dal})$ and together with $Y$'s form what
can be referred to as the large algebra. Introducing
\begin{align}
&W(Z, Y)=\go+F_{\go}(\go, C; Y,Z)+F_{\go}(\go, C, C; Y,Z)+\dots\,,\label{d1}\\
&B(Z,Y)=C+F_{C}(C, C; Y,Z)+F_{C}(C, C, C; Y,Z)+\dots\,,\label{d2}
\end{align}
and setting
\begin{align}
&\dr_x W+W\star W=0\,,\label{e1}\\
&\dr_x B+[W,B]_\star=0\,,\label{e2}
\end{align}
where $\go$ and $C$ do not depend on $Z$, while $F_\go$ and
$F_{C}$ are supposed to encode the would be HS vertices on the
right of \eqref{eq1}-\eqref{eq2}. Such an extension requires star
product $\star$ to act not only on $Y$ variables but on $Z$ as
well and as such should extend \eqref{exp} preserving
associativity. System \eqref{e1}-\eqref{e2} is obviously
consistent under $\dr_x^2=0$. Therefore, in order to have a form
of \eqref{eq1}-\eqref{eq2}, the dependence on $Z$ must identically
vanish $\dz(\eqref{e1}, \eqref{e2})\equiv 0$. The requirement of
course imposes a problem of the $Z$ -- dependence of functions
\eqref{d1}, \eqref{d2}. This problem was solved by Vasiliev
\cite{Vasiliev:1992av} as he proposed the following evolution
equations along $Z$ by introducing an auxiliary field\footnote{The
original Vasiliev equations have a slightly different form with
field $S$ instead of $\Lambda$. The two are related by a shift
$S=\Lambda+\ff{i}{2}\theta^A Z_{A}$ for the original commutation
relations of $Z$'s}. $\Lambda(Z,
Y)=\theta^{\al}\Lambda_{\al}+\bar\theta^{\dal}\bar\Lambda_{\dal}$
\begin{align}
&\dz W+\{W,\Lambda\}_{\star}+\dr_x\Lambda=0\,,\label{ee1}\\
&\dz B+[\Lambda,B]_{\star}=0\label{ee2}\,,
\end{align}
where we denote
\be
\dr z^{\al}\equiv\theta^{\al}\,,\qquad \dr\bar
z^{\dal}\equiv\bar\theta^{\dal}\,.
\ee
Now by acting with $\dr_z$ one easily verifies that $\eqref{e1}$
and \eqref{e2} are free from the $z$ -- dependence. Eqs.
\eqref{ee1}-\eqref{ee2} determine otherwise unspecified functions
$F_{\go}$ and $F_{C}$ in terms of a yet unknown connection
$\Lambda$. The missing equation for $\Lambda$ can be guessed by
checking consistency of \eqref{ee1}-\eqref{ee2}, $\dz^2=0$, that
fulfills by setting
\be\label{Seq}
\dz\Lambda+\Lambda\star\Lambda=\eta B\star\gamma+\bar\eta
B\star\bar\gamma\,,
\ee
where $\gga$ and $\bar\gga$ are the so called Klein two-forms,
which we specify in what follows and $\eta$ is a free parity
breaking parameter. The reason why the Klein two-forms $\gga$ and
$\bar\gga$ appear in \eqref{Seq} is due to the twisting in
commutators \eqref{e2} and \eqref{ee2} that we sloppily assume.
Eqs. \eqref{e1}-\eqref{e2}, \eqref{ee1}-\eqref{ee2} and
\eqref{Seq} are called the Vasiliev equations.

One notes that from the point of view of a mere consistency and a
formal form of \eqref{eq1}-\eqref{eq2}, the Vasiliev equations
allow for any star operation $\star$ so long it is associative,
admits a well-defined Klein two-form $\gga$ and leads to a regular
product of master fields. We are going to introduce this freedom
following \cite{Didenko:2019xzz} as a one parameter re-ordering of
the original star product.

\subsection{$\gb$ -- reordered form. Holomorphic sector}\label{reord}
From now on we restrict ourselves to the holomorphic sector, --
that is we set $\bar\eta=0$. The HS generating system then amounts
to
\begin{align}
&\dr_x W+W\star W=0\,,\label{h1}\\
&\dr_x B+W\star B-B\star\pi(W)=0\,,\label{h2}\\
&\dz W+\{W,\Lambda\}_{\star}+\dr_x\Lambda=0\,,\label{h3}\\
&\dz B+\Lambda\star B-B\star\pi(\Lambda)=0\label{h4}\,,\\
&\dz\Lambda+\Lambda\star\Lambda=B\star\gamma\,,\label{h5}
\end{align}
where $\star$ is the standard Vasiliev star product. Since we are
in the holomorphic sector one needs no $\bar z$ variable from
$Z=(z, \bar z)$
\be\label{star}
(f\star g)(z, Y)=\ff{1}{(2\pi)^2}\int \dr^2 u \dr^2 v f(z+u; y+u;
\bar y)\bar\star\,g(z-v; y+v; \bar y)\exp(iu_{\al}v^{\al})
\ee
and $\bar\star$ is the leftover star product acting on $\bar y$
\eqref{barexp}. We are going to systematically omit $\bar\star$ as
well as the $\bar y$ -- dependence itself from now onwards.

We use the manifest form of the twisting now
\be\label{pi}
\pi(f(y,z))=f(-y, -z)
\ee
instead of the usual outer Klein operators, which we do not
introduce in our definitions. The two-form $\gamma$ is given by
\be
\gga=\ff12e^{iz_{\al}y^{\al}}\theta_{\gb}\theta^{\gb}\,.
\ee
In what follows it is convenient to introduce the following
notation that we will use to simplify expressions. For two
commuting spinors $a_{\al}$ and $b_{\gb}$ let us define their
contraction as follows
\be\label{not}
a_{\al}b^{\al}\equiv ab=-ba\,.
\ee
Now, following \cite{Didenko:2019xzz} one can replace the original
star product \eqref{star} with the $\gb$ -- reordered one that
arises in the locality analysis
\be\label{gepstar}
f\star_{\gb} g= \int\ff{\dr^2 u\dr^2 u'\dr^2 v\dr^2 v'}{(2\pi)^4}
f(z+u', y+u)g(z-(1-\gb)v-v',y+v+(1-\gb)v')
\exp({iu_{\al}v^{\al}+iu'_{\al}v'^{\al}})\,.
\ee
The commutation relations are
\begin{align}
&y\star_\gb =y+i\ff{\p}{\p y}-i(1-\gb)\ff{\p}{\p z}\,,\qquad
\star_\gb y=y-i\ff{\p}{\p y}-i(1-\gb)\ff{\p}{\p
z}\,,\\
&z\star_\gb =z-i\ff{\p}{\p z}+i(1-\gb)\ff{\p}{\p y}\,,\qquad
\star_\gb z=z+i(1-\gb)\ff{\p}{\p y}+i\ff{\p}{\p z}\,.
\end{align}
Note that the product reduces to \eqref{exp} for $z$ --
independent functions and to \eqref{star} for $\gb=0$. One can
also observe that
\be
f(y)\star_{\gb}g(z,y)=f(y)\star g(z,y)\,,\qquad
g(z,y)\star_{\gb}f(y)=g(z,y)\star f(y)\,,
\ee
which makes $\star_{\gb}$ indistinguishable from the original
Vasiliev star product at lower orders, where no product of two $z$
-- dependent functions is yet present. Parameter $\gb$
interpolates between different orderings. So, the Weyl ordering
corresponds to $\gb=1$, while $\gb=2$ corresponds to the
anti-Vasiliev one. An operator that maps symbols in the original
ordering \eqref{star} to the re-ordered one is
\begin{align}
&O_{\gb}f(z,y)=\int\ff{\dr^2 u\dr^2 v}{(2\pi)^2} f(z+v,y+\gb u)\exp (iu_{\al}v^{\al})\,,\label{ord}\\
&O^{-1}_{\gb}f(z,y)=O_{-\gb}f(z,y)=\int\ff{\dr^2 u\dr^2
v}{(2\pi)^2}f(z+ v,y-\gb u)\exp (iu_{\al}v^{\al})\,.
\end{align}
Using it one easily calculates Klein two-form $\gga_{\gb}$
corresponding to ordering \eqref{gepstar} that appears in
\eqref{h5}, \cite{Didenko:2019xzz}
\be\label{gbga}
\gga_{\gb}=\ff12O_{\gb}(\exp({izy}))\theta_{\al}\theta^{\al}=
\ff12\ff{1}{(1-\gb)^2}\exp\left({\ff{i}{1-\gb}zy}\right)
\theta_{\al}\theta^{\al}\,.
\ee
Note, that $\gga$ remains invariant if the re-ordering is
supplemented by the proper rescaling of $z$. It makes it
convenient introducing
\be\label{Oresc}
\mathcal{O}_{\gb} f(z,y):=O_{\gb} f(z,y)\Big|_{z\to(1-\gb)z}
\ee
leading to\footnote{Note that rescaling $z\to(1-\gb)z$ implies
$\dr z=\theta\to(1-\gb)\theta$. }
\be\label{in}
\mathcal{O}_{\gb}\gga=\gga\,,\quad\forall\gb
\ee

Remarkably, as far as the lower orders are concerned, the $\gb$ --
re-ordered Vasiliev equations result in vertices that do not
depend on $\gb$ at all. These are $\Upsilon(\go, \go, C)$ and
$\Upsilon(\go, C, C)$. This fact was noted in
\cite{Didenko:2019xzz} and earlier for the case of the central
on-mass-shell theorem \eqref{omg} calculated for $\gb=1$ in
\cite{DeFilippi:2019jqq}. In particular, these vertices are
insensitive of $\gb\to-\infty$. Moreover, the perturbative master
fields $\Lambda^{(1)}$, $W^{(1)}$ and $B^{(2)}$ do not depend on
$\beta$, upon trivial rescaling $z\to (1-\gb)z$ either. This
suggests that at least at lower orders the functional class of the
Vasiliev master fields is $\gb$-ordering independent (modulo
trivial rescaling of $z$) functions
\begin{align}\label{inv}
&\mathcal{O}_{\gb}(W)=W\,,\nn\\
&\mathcal{O}_{\gb}(B)=B\,,\\
&\mathcal{O}_{\gb}(\Lambda)=\Lambda\,.\nn
\end{align}
We may emphasize that from the point out  of view of HS
consistency and lower order interaction, star product
\eqref{gepstar} is not any worse or better than the original one
\eqref{star}. What can make a difference is the functional class
that should correspond to the perturbatively well-defined
equations \eqref{eq1}-\eqref{eq2}. This class is of course
sensitive to a particular star product.

\subsection{Limit $\gb\to-\infty$}\label{lim}

While the lower order vertices are insensitive to a particular
$\gb$ -- ordering, things change at the level of $\Upsilon(\go,
\go, C, C)$ and higher, where parameter $\beta$ essentially
survives and it is only at $\gb\to-\infty$ that one recovers an
(ultra-)local result. There are different ways of thinking of what
goes on at higher orders. From the point of view of functional
class \eqref{inv} (yet to be specified), star-product \eqref{star}
does not respect its invariance as soon as both multipliers are
$z$ -- dependent. To keep fields within the class amounts to
introducing the special contracting homotopy operators for solving
for the master field $z$ -- dependence that somehow undo the
effect of star-product \eqref{star}. Technically this procedure
assumes a limit $\gb\to-\infty$, which enters the homotopy
operators. Given the effect of $\gb$ is equivalent to the
re-ordering and rescaling, $\gb\to-\infty$ looks like a certain
contraction of the original star product \eqref{star}. This
suggests that the limiting star-product respects functional class
\eqref{inv} and might be implemented at the level of the Vasiliev
equations.

Let us have a detailed look at whether such a limit really makes
sense at the level of equations of motion. Naively,
$\gb\to-\infty$ can be straightforwardly taken from
\eqref{gepstar}. To carry it out one should keep in mind to
rescale properly (see also \cite{Didenko:2019xzz})
\be\label{rsc}
z\to (1-\gb)z\,,\qquad \theta=\dr z\to (1-\gb)\theta\,.
\ee
This amounts to set the multiplied functions to be
$f=f(\ff{z}{1-\gb}, y)$. The result for the limiting star product
$\gb\to-\infty$ gets immediately available then,
$*\equiv\star_{-\infty}$
\be\label{limst}
f*g= \int\ff{\dr^2 u\dr^2 u'\dr^2 v\dr^2 v'}{(2\pi)^4}
f\left(z+u', y+u\right)g\left(z-v,y+v+v'\right)
\exp({iu_{\al}v^{\al}+iu'_{\al}v'^{\al}})\,.
\ee
The obtained $*$ -- product is naturally associative and gives the
following rules
\begin{align}
&y* =y+i\ff{\p}{\p y}-i\ff{\p}{\p z}\,,\qquad z* =z+i\ff{\p}{\p
y}\,,\\
&* y=y-i\ff{\p}{\p y}-i\ff{\p}{\p z}\,,\qquad * z=z+i\ff{\p}{\p
y}\,.
\end{align}
Some important properties of $*$ are in order. Unlike the original
product \eqref{star}, oscillators $z$ commute in the limiting case
\be
[z_{\al}, z_{\gb}]_{*}=0\,.
\ee
This means that limit $\gb\to-\infty$ is really a contraction
rather than just a re-ordering. Another feature is whenever one of
the multiplier is $z$ -- independent, \eqref{limst} acts exactly
as \eqref{star}, e.g.,
\be\label{equiv}
f(y)*g(z,y)=f(y)\star g(z,y)\,,\quad g(z,y)*f(y)=g(z,y)\star f(y)
\ee
going in line with a more general statement of $\gb$ --
independence of lower order vertices.

Eqs. \eqref{h1}-\eqref{h5} formally survive limit $\gb\to-\infty$
and preserve the same form with the natural replacement $\star \to
*$. In taking the limit the form of the Vasiliev equations in
terms of field $\Lambda$ rather than $S$ (see footnote 9) was
important. The naive limit just implemented is not harmless
though. The room for a potential problem is easy to isolate.
Unlike the original Klein operator, the one corresponding to
\eqref{limst} has certain ill-defined properties. For example,
\be\label{inf}
e^{izy}*e^{izy}=\infty=\gd^{2}(0)\,.
\ee
The worrisome divergency calls for a thorough analysis of the
functional class that survives under \eqref{limst}.

\subsection{Classes of functions}\label{classes}

In order to proceed with the functional class  of the contracted
Vasiliev equations analysis let us sort out their form. There are
two space-time equations
\begin{align}
&\dr_{x} W+W* W=0\,,\label{g1}\\
&\dr_{x} B+W* B-B*\pi(W)=0\,,\label{g2}
\end{align}
two equations that determine field $z$ -- dependence
\begin{align}
&\dz W+\{W,\Lambda\}_{*}+\dr_x\Lambda=0\,,\label{g3}\\
&\dz B+\Lambda* B-B*\pi(\Lambda)=0\label{g4}
\end{align}
and an equation for the auxiliary field $\Lambda$
\be
\dz\Lambda+\Lambda*\Lambda=B*\gamma\,,\qquad
\gga=\ff12e^{izy}\theta\theta\,.\label{g5}
\ee
To identify the required class of functions one notes that fields
entering the equations are $\theta$ -- graded. $W$ and $B$ are
zero-forms in $\theta$, while $\Lambda$ is a one-form and $\gga$
is a two-form. The class of functions we look for
${\mathbf{C}}^r=\{\phi(z,y; \theta)\}$ can be labelled by the
$\theta$ -- degree $r=0,1,2$ for zero-, one- and two-forms,
correspondingly. We require ${\mathbf{C}}^r$ to be closed under
$*$ and respect $\theta$ -- grading
\begin{align}
&{\mathbf{C}}^{r_1}*{\mathbf{C}}^{r_2}\to{\mathbf{C}}^{r_1+r_2}\,,\label{cl1}\\
&\dz {\mathbf{C}}^{r}\to{\mathbf{C}}^{r+1}\,.\label{cl2}
\end{align}
Note, that since $\theta\theta\theta=0$ this implies
${\mathbf{C}}^{3}=\O$. In addition to \eqref{cl1}-\eqref{cl2}, our
definition of the class includes invariance \eqref{inv} under
re-ordering \eqref{Oresc}, i.e.,
\be\label{cl3}
{\mathbf{C}}^r=\{\phi(z,y; \theta):\quad
\mathcal{O}_{\gb}(\phi)=\phi\,,\quad \forall\phi_{1,2}:\quad
\mathcal{O}_{\gb}(\phi_1*\phi_2)=\phi_1*\phi_2\}\,.
\ee
While we do not have a clear insight into condition \eqref{inv},
we may note that it holds experimentally at a few available
interaction orders, as we have stressed earlier. So, our strategy
is to look at it as the all order exact. Condition \eqref{cl3} may
happen to be in tension with \eqref{cl1}, \eqref{cl2}, but it
turns out that the all three are consistent if well-defined with
star product \eqref{limst}, unlike \eqref{star}. The details of
derivation of ${\mathbf{C}}^r$ are given in the Appendix A. The
final result is a subclass of the one proposed earlier in
\cite{Gelfond:2019tac}. It includes functions of the following
form
\be\label{class}
\phi(z, y; \theta)=\int_{0}^{1}\dr\tau\ff{1-\tau}{\tau}\int
\ff{\dr u\,\dr v}{(2\pi)^2}\, f\Big(\tau z+v, (1-\tau)(y+u);
\ff{\tau}{1-\tau}\theta\Big)e^{i\tau
z_{\al}y^{\al}+iu_{\al}v^{\al}}\,,
\ee
where $f$ is such that integration over $\tau$ makes sense and is
otherwise arbitrary. A convenient way of looking at \eqref{class}
is using the generating functions. Taking
$f\sim\exp{(iyA+izB+i\gep A\cdot B)}$ with sources $A$ and $B$ and
conveniently setting $\gep=1$ for $r=0$ and $\gep=0$ for $r=1,2$
in order to control over $\tau$ -- poles one then finds by
explicit integration over $u$ and $v$ in \eqref{class}
\begin{align}
&\theta^0:\qquad\int_{0}^{1}\dr\tau\,\ff{1-\tau}{\tau}e^{i\tau
z_{\al}y^{\al}+i(1-\tau)A^{\al}y_{\al}+i\tau B^{\al}z_{\al}-i\tau
A^{\al}B_{\al}}\,,\label{t0}\\
&\theta^1:\qquad\int_{0}^{1}\dr\tau\, e^{i\tau
z_{\al}y^{\al}+i(1-\tau)A^{\al}y_{\al}+i\tau
B^{\al}z_{\al}+i(1-\tau) A^{\al}B_{\al}}\,,\label{t1}\\
&\theta^2:\qquad\int_{0}^{1}\dr\tau\,\ff{\tau}{1-\tau}e^{i\tau
z_{\al}y^{\al}+i(1-\tau)A^{\al}y_{\al}+i\tau
B^{\al}z_{\al}+i(1-\tau) A^{\al}B_{\al}}\,.\label{t2}
\end{align}
Now, any function $\phi$ from \eqref{class} can be viewed as being
originated from \eqref{t0}-\eqref{t2} by means of the
decomposition in sources $A$, $B$. For example, at $r=0$ the
generating function should be at least linear in $B$ in order to
be well-defined as an integral over $\tau$. Similarly, it should
be at least linear in $A$ for $r=2$.

 Let us note a persistent factor $\exp{(i\tau zy)}$
as well as a specific $\tau$ -- dependence of $z$, $y$ and
$\theta$ that class \eqref{class} has. These are characteristic
features of the functional class of \cite{Vasiliev:2015wma},
\cite{Gelfond:2019tac} too. What makes \eqref{class} different is
a specific 'tail behavior' of the $AB$ -- contractions in
\eqref{t0}-\eqref{t2}. It is these contractions that are necessary
for \eqref{class} to be $\mathcal{O}_{\gb}$ -- invariant,
\eqref{Oresc}. Moreover, the original star product \eqref{star}
does not respect the invariance. In other words, $\star$ --
product of two functions from \eqref{class} does not remain in the
class in general.

A convenient way to operate with \eqref{t0}-\eqref{t2} is to note
that these can be rewritten in a somewhat factorized form
\be\label{hf}
e^{i\tau z_{\al}y^{\al}+i(1-\tau)A^{\al}y_{\al}+i\tau
B^{\al}z_{\al}-i\tau A^{\al}B_{\al}}=e^{iA^{\al}y_{\al}}\hf
e^{i\tau z_{\al}(y+B)^{\al}}\,,
\ee
where
\be\label{half}
f(y)\hf g(z,y)= \int\ff{\dr u\dr v}{(2\pi)^2} f( y+u)g(z-v,y)
\exp({iu_{\al}v^{\al}})
\ee
is a certain product. Note that the left multiplier in
\eqref{half} is $z$ -- independent. We are going to use the
factorized representation \eqref{hf} throughout the paper in what
follows.

Let us look closer at \eqref{class}. As HS equations
\eqref{eq1}-\eqref{eq2} are formulated in terms of $z$ --
independent functions the natural question is whether
\eqref{class} contains functions of variable $y$ only. The answer
to this question is affirmative within the $r=0$ class.

\paragraph{$z$ -- independent functions}

Consider the ${\mathbf{C}}^0$ sector and let us pick the following
$\phi(z,y)\in {\mathbf{C}}^0$
\be
\phi(z,y)=\int_{0}^{1}\dr\tau f^{\al}(y)\hf(1-\tau)
z_{\al}e^{i\tau z(y+q)}\equiv
-e^{izy}\p^{y}_{\al}\int_{0}^{1}\dr\tau\left(f^{\al}(x)e^{-izx}\right)\,,
\ee
where $f^{\al}(y)$ is yet an arbitrary function and $q$ is an
arbitrary spinor (it can be zero, for example) and
\be
x=(1-\tau)y-\tau q\,.
\ee
Choosing now $f_{\al}=-(y+q)_{\al}f(y)$ we can easily see that
\be
\phi(z,y)=f(y)\,,
\ee
i.e., $\phi$ is $z$ -- independent. Indeed,
\begin{align}
&-e^{izy}\p^{y}_{\al}\int_{0}^{1}\dr\tau
f^{\al}(x)e^{-izx}=e^{izy}\p^{y}_{\al}\int (1-\tau)
(y+q)^{\al}f(x)e^{-izx}=\\
&=e^{izy}\int_{0}^{1}\dr\tau\left(2(1-\tau)-(1-\tau)^2\ff{\p}{\p
\tau}\right)f(x)e^{-izx}=-e^{izy}\int_{0}^{1}\dr\tau
\p_{\tau}\left((1-\tau)^2f(x)e^{-izx}\right)=f(y)\,,\nn
\end{align}
where we integrated by parts over $\tau$ in the last line. There
are many ways to represent one and the same $z$ -- independent
function. For example, up to a coefficient it can be represented
as follows
\be
f(y)\sim \int_{0}^{1}\dr\tau
(y+q)^{\al(n)}f(y)\hf(1-\tau)\tau^{n-1} z_{\al}\dots z_{\al}
e^{i\tau z(y+q)}
\ee
as can be checked by the multiple partial integration. Therefore,
any $z$ -- independent $\theta$ -- zero-form belongs to
${\mathbf{C}}^0$.

\paragraph{Klein operator}
Another important object that manifests in \eqref{g5} is Klein
operator $\gk=e^{izy}$. Notably, there is no such function within
${\mathbf{C}}^0$. Instead, it resides in sector ${\mathbf{C}}^2$
just as it appears in \eqref{g5} as a two-form $\gga$. To find it
consider the following element from ${\mathbf{C}}^2$
\be
\gk:=\int_{0}^{1}\dr\tau\ff{\tau}{1-\tau}\,\gep^{\al\gb}\ff{\p^2}{\p
A^{\al} \p B^{\gb}}\,e^{i\tau
z_{\al}y^{\al}+i(1-\tau)A^{\al}y_{\al}+i\tau
B^{\al}z_{\al}+i(1-\tau) A^{\al}B_{\al}}\Big|_{A=B=0}\,.
\ee
One then finds by partial integration,
\be
\gk=\int_{0}^{1}\dr\tau (2\tau+i\tau^2 zy)\,e^{i\tau
zy}=e^{izy}\,.
\ee

\subsubsection{Star products}\label{sec}
Class \eqref{cl1}-\eqref{cl3} seemingly contains all the required
elements to operate on \eqref{g1}-\eqref{g5}. The important
question however is whether star product \eqref{limst} provides a
meaningful outcome. We noted already that the replacing of the
original product \eqref{star} with \eqref{limst} may not be a
harmless endeavor, \eqref{inf}. So, let us check out if the left
hand side of \eqref{cl1} really makes sense. The good news is
\be\label{prw}
({\mathbf{C}}^{0}*{\mathbf{C}}^{0}\,,\quad
{\mathbf{C}}^{0}*{\mathbf{C}}^{1}\,,\quad
{\mathbf{C}}^{1}*{\mathbf{C}}^{0})-{\textnormal{well-defined}}
\ee
This can be checked straightforwardly by using \eqref{t0} and
\eqref{t1} as one finds that the $\tau$ -- integrations result in
analytic expressions in that case. The problem blows up for other
products
\be\label{pri}
({\mathbf{C}}^{1}*{\mathbf{C}}^{1}\,,\quad
{\mathbf{C}}^{0}*{\mathbf{C}}^{2}\,,\quad
{\mathbf{C}}^{2}*{\mathbf{C}}^{0})-{\textnormal{ill-defined}}\,,
\ee
unless ${\mathbf{C}}^{0}$ is $z$ -- independent in which case
${\mathbf{C}}^{0}*{\mathbf{C}}^{2}$ and
${\mathbf{C}}^{2}*{\mathbf{C}}^{0}$ exist and coincide with those
calculated using the Vasiliev star product \eqref{equiv}.

As an instructive example let us look at the product of $f\in
{\mathbf{C}}^{0}$ given by
\be\label{ex1}
f_{\al}=\int_{0}^{1}\dr\tau\,(1-\tau)z_{\al}e^{i\tau zy}
\ee
with the Klein. From \eqref{limst} one has
\be
f(z,y)*e^{izy}=\int\ff{\dr u\dr v}{(2\pi)^2}\,
e^{iz_{\al}(y+v)^{\al}+iu_{\al}v^{\al}}f(v, u)
\ee
and thus substituting \eqref{ex1} we find
\be\label{inf02}
f_{\al}*e^{izy}=e^{izy}\int_{0}^{1}\dr\tau\ff{1}{1-\tau}\int\dr
v\gd(v)\,v_{\al}=\infty\cdot 0\,.
\ee
Uncertainty \eqref{inf02} illustrates a problem with star product
\eqref{limst} for the case of ${\mathbf{C}}^{0}*{\mathbf{C}}^{2}$
and ${\mathbf{C}}^{2}*{\mathbf{C}}^{0}$. It does not exist unless
$\dr_z {\mathbf{C}}^{0}=0$. In other words, the product makes
sense for $z$ -- independent functions $f(y)\in {\mathbf{C}}^{0}$
only.

Similarly, one can look at ${\mathbf{C}}^{1}*{\mathbf{C}}^{1}$.
Choosing, for example,
\be
\Lambda_{\al}=\int_{0}^{1}\dr\tau\,\tau z_{\al}e^{i\tau z(y+p)}\,,
\ee
where $p$ is an arbitrary spinor parameter, we analogously arrive
at
\be\label{inf11}
\Lambda_{\al}*\Lambda^{\al}=\infty\cdot 0\,.
\ee
This example illustrates a general phenomenon that star product
\eqref{limst} is ill-defined for certain products that however
inevitably present perturbatively within HS generating equations
\eqref{g1}-\eqref{g5}. More precisely, star product \eqref{limst}
is consistent with most of the generating equations
\eqref{g1}-\eqref{g4}, because these equations contain products of
type \eqref{prw}. However, it does not allow one to determine
evolution of field $\Lambda$ from \eqref{g5} beyond free level. At
first order $\Lambda^{(1)}$ can be determined from \eqref{g5}
because the right hand side contains $C(y)*\gga$ which is well
defined being a product of a $z$ -- independent function by a
two-form.

A conclusion here is eqs. \eqref{g1}-\eqref{g5} can not be taken
as the generating ones for higher-spin system
\eqref{eq1}-\eqref{eq2} beyond lower orders. Precisely, they
reproduce $\Upsilon(\go,\go, C)$ and $\Upsilon(\go,C, C)$ and then
stubborn in $0\cdot\infty$ at higher orders. Note the stark
contrast with the original Vasiliev product \eqref{star}, which
has no obstruction in formal $\Upsilon$ reconstruction at any
orders. One arrives at a curious case. The analysis of
\cite{Didenko:2019xzz} plainly indicates the emergence of algebra
\eqref{limst} via the special homotopy resolutions as soon as one
insists on higher order locality. Yet, a naive re-ordering
followed by a contraction $\gb\to-\infty$ at the level of the
Vasiliev equations that leads to \eqref{limst} seemingly makes no
sense in \eqref{g5}. The situation gets even more mysterious if
noted that vertex $\Upsilon_{C\go\go C}$ residing in $W_1\star
W_1$ part of \eqref{e1} was effectively calculated using $*$ in
place of $\star$ in \cite{Didenko:2019xzz}.

A possible resolution is that the Vasiliev equations may admit a
consistent modification that keeps space-time equations \eqref{e1}
and \eqref{ee1} intact, while featuring a different condition for
$\Lambda$ in the case of product \eqref{limst}.

\section{Generating equations for the holomorphic sector}\label{gen}

Let us start over with the derivation of the Vasiliev type
generating equations. We keep the original idea that \eqref{eq1}
can be reached by setting
\begin{align}
&\dr_{x} W+W*W=0\,,\label{dxw}\\
&\dz W+\{W,\Lambda\}_{*}+\dr_x\Lambda=0\,,\label{dzw}
\end{align}
where we replace Vasiliev product \eqref{star} with \eqref{limst}.
We also constrain ourselves to the specific functional class
\eqref{class}
\be
W\in {\mathbf{C}}^{0}\,,\quad \Lambda\in {\mathbf{C}}^{1}\,,
\ee
which makes all operations within \eqref{dxw} and \eqref{dzw}
well-defined. Recall, condition \eqref{dzw} guarantees that
\eqref{dxw} is $z$ -- independent as can be checked by applying
$\dr_{z}$ to \eqref{dxw} and making use of \eqref{dzw}. In so
doing one never encounters undefined operations \eqref{pri}.
Similarly, applying $\dr_x$ to \eqref{dzw} one finds no further
constraints or ill-defined structures either. A pinnacle of the
problem is of course the $\dr_z$ -- consistency of \eqref{dzw}
that normally leads to the introduction of the zero-form module
$B$, \eqref{e2} and eventually to \eqref{g5}. In our case,
however, \eqref{g5} makes no sense as it contains structures
${\mathbf{C}}^{1}*{\mathbf{C}}^{1}$ and
${\mathbf{C}}^{0}*{\mathbf{C}}^{2}$. To check whether \eqref{dzw}
may have solutions or not, we apply $\dr_z$ to see that there are
none unless
\be\label{const1}
\dr_z\{W,\Lambda\}_{*}=\dr_x\dr_z\Lambda\,.
\ee
The important comment is while both parts of \eqref{const1} are
well-defined on classes \eqref{class}, neither $[\dr_z W,
\Lambda]_*$ nor $[W, \dr_z\Lambda]_*$ being products of type
\eqref{pri} exists, unless $W$ is in cohomology of $\dr_z$, i.e.,
$z$ -- independent, \eqref{equiv}. It is this feature that
prevents one from re-expressing $\dr_z W$ from \eqref{dzw} for
substitution into \eqref{const1}. Equivalently, one can not use
the Leibniz rule on the left hand side of \eqref{const1}.
Specifically, if one takes e.g. $\dr_z W$ from \eqref{dzw} and
plug it into $[\dr_z W, \Lambda]_*$ there would be the terms
containing products of two $\Lambda$'s. These however are
ill-defined being of the type ${\mathbf{C}}^{1}*{\mathbf{C}}^{1}$.
Let us stress once again that the original expression
$\dr_z\{W,\Lambda\}_{*}$ is perfectly alright. Indeed, once
$\{W,\Lambda\}_{*}$ is well defined so is
$\dr_z\{W,\Lambda\}_{*}$\footnote{As an illustration, one can
think of $\dr_z ({\mathbf{C}}^{0}*{\mathbf{C}}^{1})=\dr_z
{\mathbf{C}}^{0}*{\mathbf{C}}^{1}+
{\mathbf{C}}^{0}*\dr_z{\mathbf{C}}^{1}$ as of a decomposition of a
convergent integral into a sum of two divergent ones.}.

To proceed further we need an equation that determines evolution
of $\Lambda$ along $z$. Typically of the Vasiliev-like approach,
such an evolution is determined in terms of $\Lambda$ itself and
in terms of some new zero-form field $B(z,y)\in {\mathbf{C}}^{0}$.
As a matter of principle all possible contributions to a
$\dr_z\Lambda$-equation can contain the two-form $\Lambda*\Lambda$
and a two-form composed of $B$. The latter being a zero-form
should be accompanied with a certain two-form
$\Gamma\in{\mathbf{C}}^{2}$. The problem here is that once
$\dr_z\Lambda\in {\mathbf{C}}^{2}$, it can not be expressed via
undefined ${\mathbf{C}}^{1}*{\mathbf{C}}^{1}$. This excludes the
$\Lambda*\Lambda$ contribution. Similarly,
${\mathbf{C}}^{2}*{\mathbf{C}}^{0}$ and
${\mathbf{C}}^{0}*{\mathbf{C}}^{2}$ are both ill-defined unless
the ${\mathbf{C}}^{0}$-multiplier is $z$ -- independent implying
that $B$ can not depend on $z$, $B:=C(y)$. Therefore, the
remaining option is a product of $C$ by
$\Gamma\in{\mathbf{C}}^{2}$. At free level such dependence is
given by
\be\label{const2}
\dr_z\Lambda=C*\gga\,,
\ee
where $C(y)$ can depend on $y$ only and $\Gamma=\gga$. Let us
stress that any $z$ -- dependent corrections of the form
\eqref{t0} to field $C$ will lead to a meaningless result. This
implies that \eqref{const2} should be taken as either all-order
exact or the higher order corrections modify\footnote{We thank the
anonymous Referee for pointing out this option to us.}
$\gga\to\gga^{int}=\gga+O(C)$ (modulo field redefinition
$\gga^{int}\to f(C)*\gga$). While the latter alternative can not
be {\it a priori} excluded, we will see that \eqref{const2} can
indeed be taken as the exact one upon specifying its solution. So,
our strategy is to postulate \eqref{const2} in what follows.
Plugging then \eqref{const2} into \eqref{const1} one arrives at
\be\label{dxc}
\dr_x C*\gga=\dr_z\{W,\Lambda\}_{*}\,,
\ee
which leads to a yet another consistency check with respect to
$\dr_x^2=0$. As the left hand side is $\dr_x$ -- exact, the right
one should satisfy $\dr_z\dr_x\{W,\Lambda\}_{*}=0$, which is
indeed the case in view of \eqref{dxw} and \eqref{dzw}. As a
result, eqs. \eqref{dxw} and \eqref{dzw} are consistent provided
\eqref{dxc} is satisfied. The latter equation being interpreted as
\eqref{eq2} (modulo $\gga$) places a stringent constraint on the
right hand side of \eqref{dxc}. Namely, once $\dr_x C*\gga$ is a
product of $z$ -- independent function $\dr_x C(y|x)$ by Klein
two-form $\gga$, the same dependence of type ${f}(y)*\gga$ should
be on the right hand side of \eqref{dxc} too. Specifically,
comparing \eqref{dxc} with \eqref{eq2}, we have
\be\label{cgga}
\dr_x C*\gga=\left(\Upsilon(\go, C)+\Upsilon(\go, C,
C)+\Upsilon(\go, C, C, C)+\dots\right)*\gga\,,
\ee
with $\Upsilon$'s being $z$ -- independent by definition
\eqref{eq2} and where
\be\label{verres}
\Upsilon(\go, \underbrace{C,\dots,
C}_n)*\gga=\dr_z\{W^{(n-1)},\Lambda\}_{*}
\ee
and $n$ -- is the order of perturbative expansion in $C$. If this
is not the case, then \eqref{dxc} can not have the form of
\eqref{cgga}.

Let us note that at this stage our consistency analysis is general
and applicable to any associative star product and any functional
class, provided expressions make sense. In particular, one could
have written down the same system with the original star-product
\eqref{star}. Not surprisingly, \eqref{dxc} acquires a wrong
dependence on $z$ that is $C$ can not be $z$ -- independent in
this case, such that it loses interpretation in terms of
\eqref{eq2} beyond free level unless higher order corrections of
$\Lambda$ are taken into account. Therefore, it can not describe
higher-spin dynamics with \eqref{const2}. Things change radically
with star product \eqref{limst}.

\paragraph{Projector identity}
A crucial observation is -- for any function $f\in
{\mathbf{C}}^{0}$ the following projector identity takes place
\be\label{awe}
\boxed{\dr_z (f*\Lambda_0)=F_R(y)*\gga\,,\quad \dr_z
(\Lambda_0*f)=F_L(y)*\gga\,,}
\ee
where $\Lambda_0$ is the following solution of \eqref{const2}
\be\label{L0}
\Lambda_0=\theta^{\al}\int_{0}^{1}\dr\tau\,\tau z_{\al}C(-\tau
z)e^{i\tau zy}\,,
\ee
$f$ is given by \eqref{class} at $\theta=0$ and $F_{R,L}$ are the
following $z$ -- independent functions
\be\label{FR}
F_L:=(C*_{y} f(-z, -y))\Big|_{z=y}\,,\qquad F_R:=(f(-z, y)*_{y}
C)\Big|_{z=y}\,,
\ee
where $*_{y}$ is the standard Weyl star-product \eqref{exp} that
acts on variables $y$ and ignores $z$. Eq. \eqref{awe} says that
the $z$ dependence of its left hand side collapses into the Klein
two form no matter what function $f(z,y)\in {\mathbf{C}}^{0}$ is.
This result is not at all obvious and can be seen after the
detailed calculation of \eqref{awe} that boils down to a total
derivative in one of the two integrals over $\tau$'s followed by a
proper integration variable change. The projector identity plays a
central role in the HS interpretation of eq. \eqref{dxc} and we
prove it in Appendix B. That $f\in {\mathbf{C}}^{0}$ and
$\Lambda_0$ is \eqref{L0} is crucially important. In particular,
\eqref{awe} is not going to hold for the $z$ -- dependent $f$ if
one to replace $\Lambda_0$ with $\Lambda_0+\dr_z\xi$, where
$\xi\in {\mathbf{C}}^{0}$ is an arbitrary function\footnote{The
case is somewhat analogous to the early formulation of HS
equations \cite{Vasiliev:1990en}, where retrospectively the
auxiliary connection $\Lambda$ was suitably fixed. We thank M.A.
Vasiliev for the related discussion.}. It should be noted also
that \eqref{FR} are well-defined for a well-defined $f\in
{\mathbf{C}}^{0}$. Therefore, vertices $\Upsilon$ in
\eqref{verres} are indeed $z$ -- independent, while the
consistency requirement \eqref{dxc} for \eqref{dzw} to admit
solutions is fulfilled. Eventually, the complete system that
generates the holomorphic higher-spin interactions reads
\begin{align}
&\dr_{x} W+W*W=0\,,\label{dxWeq}\\
&\dz W+\{W,\Lambda_0\}_{*}+\dr_x\Lambda_0=0\,,\label{Weq}\\
&\dr_z\Lambda_0=C*\gga\,,\label{Lambdaeq}\\
&\dr_x C*\gga=\dr_z\{W,\Lambda_0\}_{*}\,,\label{Ceq}
\end{align}
where $\Lambda_0$ is given by \eqref{L0}. Let us summarize its
main features.
\begin{itemize}
\item The system makes sense as the higher-spin generating if
$W\in {\mathbf{C}}^{0}$ only, which will be shown to take place at
least in perturbations. Outside that class either some star
products are ill-defined or $C$ can not be $z$ -- independent.
Most spectacular is a peculiar $z$ -- dependence via Klein form
$\gga$ that shows up on the right hand side \eqref{Ceq} for any
$W\in {\mathbf{C}}^{0}$. This makes $\Lambda_0$ from \eqref{L0} a
unique object playing a distinguished role in the whole
construction. Recall also that ${\mathbf{C}}^{0}$ contains all $z$
-- independent functions.

 \item As different from the standard Vasiliev system \eqref{h1}-\eqref{h5},
where \eqref{eq2} is reached via the $B$ - module, eq.
\eqref{Lambdaeq} is so restrictive that constrains the zero-form
vertices in \eqref{Ceq} leaving no room for any $C$ -- corrections
typically stored in $B$. This happens because the right hand side
of \eqref{Lambdaeq} is the only meaningful ${\mathbf{C}}^{2}$ --
expression that one can write down modulo field redefinition $C\to
f(C)$, provided the Klein two-form $\gga$ is kept field
independent.

Another comment is while the $z$ -- independence of vertices
\eqref{eq2} follows from the $z$ -- independence of \eqref{h2}
within the Vasiliev framework, in our case we should set $C$ that
appears in \eqref{Lambdaeq} and \eqref{Ceq} to be $z$ --
independent in the first place. This choice might have been
inconsistent with the actual dynamics governed by the equations,
but turns out to be perfectly fine with it due to \eqref{awe}.

\item The local gauge symmetry of \eqref{dxWeq}-\eqref{Ceq}
generated by $\gep\in {\mathbf{C}}^{0} $ is easy to identify
\begin{align}
&\gd_{\gep}\Lambda_0=\dr_z\gep+[\Lambda_0,\gep]_*\,,\\
&\gd_{\gep}W=\dr_x\gep+[W, \gep]_*\,,\\
&\gga*\gd_{\gep}C=\dr_z[\gep, \Lambda_0]_*\,.\label{deltaC}
\end{align}
Note, however, the presence of $\Lambda_0$, \eqref{L0} in the
gauge transformations. While formally the invariance holds for any
$\Lambda$ satisfying \eqref{Lambdaeq}, it is only for $\Lambda_0$
that $\gd_\gep C$ from \eqref{deltaC} remains $z$ -- independent
due to \eqref{awe}.

\item The natural and simplest vacuum of \eqref{dxWeq}-\eqref{Ceq}
is the $AdS$ space-time
\begin{align}
&W_{0}=-\ff{i}{4}(\go^{\al\gb}y_{\al}y_{\gb}+\bar{\go}^{\dal\dgb}\bar{y}_{\dal}\bar{y}_{\dgb}+
2\e^{\al\dgb}y_{\al}\bar{y}_{\dgb})\,,\\
&C_0=0\,,
\end{align}
where $\go$, $\bar\go$ and $\e$ are the $AdS$ connection fields.
As expected, the Minkowski space is not a solution of the theory.
However, one can consider a contraction of the original
star-product algebra \eqref{HS}
\be
\bar y\to \rho^{-1}\bar y\,,\qquad \rho\to 0
\ee
that results in
\be\label{HScont}
[y_{\al}, y_{\gb}]_\star=2i\gep_{\al\gb}\,,\qquad [y_{\al}, \bar
y_{\dgb}]_\star=0\,,\qquad [\bar y_{\dal}, \bar
y_{\dgb}]_\star=0\,.
\ee
The contraction is well-defined on eqs. \eqref{dxWeq}-\eqref{Ceq}
and leads to a specific 'flat' limit with the vacuum of the form
\be\label{vacfl}
W_0'=-\ff{i}{4}(\go^{\al\gb}y_{\al}y_{\gb}+
\e^{\al\dgb}y_{\al}\bar{y}_{\dgb})\,.
\ee
Note, this vacuum can not be regarded as the Minkowski one since
it lacks the anti-chiral part of the Lorentz connection
$\bar\go_{\dal\dgb}$ (see also footnote 5). Put it differently,
one can choose a coordinate system on the Minkowski space-time
with $\bar\go_{\dal\dgb}=0$ which is consistent with
\eqref{vacfl}. In this case, however, the manifest Lorentz
covariance appears to be lost.

\item Finally, since $W$, $\Lambda_0$ and $C$ depend on variable
$\bar y$, which we systematically omit, one should remember of
implicit $\bar\star$ -- product present in
\eqref{dxWeq}-\eqref{Ceq}.
\end{itemize}

\subsection{Perturbation theory}\label{pert}
The higher-spin vertices in \eqref{eq1}-\eqref{eq2} can be
determined from \eqref{dxWeq} and \eqref{Ceq} order by order
\begin{align}
&\Upsilon(\go,\go, \underbrace{C,\dots,
C}_n)=-\sum_{i=1}^n\dr_x W^{(i)}-\sum_{i+j=n}W^{(i)}*W^{(j)}\,,\label{verww}\\
&\Upsilon(\go, \underbrace{C,\dots,
C}_n)*\gga=\dr_z\{W^{(n-1)},\Lambda_0\}_{*}\,.\label{Cver}
\end{align}
Note that $\dr_x W^{(i)}$ contributes to the $n^{th}$-order
\eqref{verww} for all $1\leq i\leq n$. To calculate the vertices
one should solve for $W^{(i)}$ using \eqref{Weq}. So, at
$0^{\,th}$- order in $C$ we have
\be\label{W0}
\dr_z W^{(0)}=0\quad\Rightarrow\quad W^{(0)}=\go(y)\in
{\mathbf{C}}^{0}\,.
\ee
Note that the solution belongs to ${\mathbf{C}}^{0}$ as required.
In fact, one can prove that any order solution of \eqref{Weq}
belongs to ${\mathbf{C}}^{0}$ once $\Lambda_0\in
{\mathbf{C}}^{1}$. Indeed, let $W^{(n-1)}\in {\mathbf{C}}^{0}$,
then from \eqref{cl1} it follows that $\{W^{(n-1)},
\Lambda_0\}_*\in {\mathbf{C}}^{1}$ and therefore
\be
\dr_z W^{(n)}=X=\theta^{\al}X_{\al}\,,\qquad
X\in{\mathbf{C}}^{1}\,,
\ee
where $X$ comes from the $(n-1)$-order solution and by
construction belongs to ${\mathbf{C}}^{1}$
\be
X_{\al}=\int_{0}^{1}\dr\tau\int \ff{\dr u\,\dr v}{(2\pi)^2}\,
f^{(n-1)}_{\al}\Big(\tau z+v, (1-\tau)(y+u)\Big)e^{i\tau
zy+iuv}\,.
\ee
Solving for $W^{(n)}$ using the standard homotopy
\be\label{stnd}
\Delta_0
X:=z^{\al}\ff{\p}{\p\theta^{\al}}\int_{0}^{1}\dr\tau\ff{1}{\tau}
X(\tau z, y; \tau\theta)
\ee
and carry out a simple integration variable change one arrives at
\be\label{Wgen}
W^{(n)}=\int_{[0,1]^2}\dr\tau\dr\gs\int \ff{\dr u\,\dr
v}{(2\pi)^2}\,\ff{1-\tau}{\tau}\,\tau z^{\al}
f^{(n-1)}_{\al}\Big(\tau z+v, \gs(1-\tau)( y+u)\Big)e^{i\tau
zy+iuv}+\phi^{(n)}(y)\,.
\ee
The result is by definition \eqref{class} belongs to
${\mathbf{C}}^{0}$. Indeed, the first term in \eqref{Wgen} is
manifestly in ${\mathbf{C}}^{0}$ and since the freedom in the
homogeneous solution $\phi$ is any $z$ -- independent function
which too belongs to ${\mathbf{C}}^{0}$, we conclude that any
perturbative $W^{(n)}\in {\mathbf{C}}^{0}$. Although not
necessary, a convenient choice is to set $\phi^{(n)}=0$.

\section{Vertices}\label{vert}
In what follows it is convenient to use the notation from
\cite{Didenko:2018fgx}. We switch to the Taylor form of our fields
by introducing
\begin{align}
&C(y, \bar y)\equiv e^{-ip^{\al}y_{\al}}C(y',\bar
y)\Big|_{y'=0}\,,\quad p_{\al}=-i\ff{\p}{\p y^{'\al}}\,,\label{conv1}\\
&\go(y, \bar y)\equiv e^{-it^{\al}y_{\al}}\go(y',\bar
y)\Big|_{y'=0}\,,\quad t_{\al}=-i\ff{\p}{\p
y^{'\al}}\,,\label{conv2}
\end{align}
where we intentionally tagged the same operator with either $p$ or
$t$ to distinguish its action on $C$ from its action on $\go$. In
this terms any vertex that system \eqref{dxWeq}-\eqref{Ceq}
produces has the following schematic form
\begin{align}
&\Upsilon_1=\Phi^{[\delta_1, \delta_2]}(y; t_1, t_2,
p_i)\left(C\bar{\star}\dots\bar{\star}\,\go\bar{\star}\dots\bar{\star}\,
\go\bar{\star}\dots\bar{\star}\,C\right)(y'_{I}, \bar y)\Big|_{y'_{I}=0} \,,\label{ups1}\\
&\Upsilon_0=\Phi^{[\delta]}(y; t,
p_j)\left(C\bar{\star}\dots\bar{\star}\,\go\bar{\star}\dots\bar{\star}\,C\right)
(y'_{J}, \bar y)\Big|_{y'_{J}=0}\,,\label{ups0}
\end{align}
where $\Upsilon_1$ and $\Upsilon_0$ contribute to \eqref{eq1} and
\eqref{eq2} correspondingly. All $C$'s on the right are ordered as
they stand from left to right with one or two ordered $\go$ --
'impurities' that may appear in any place of the string. Each $p$
acts on the corresponding $C$ and each $t$ acts on its $\go$ and
$\#I=\#i+2$, $\#J=\#j+1$. The structure of vertices is therefore
totally encoded by functions $\Phi$'s, which however depend on the
place of $\go$'s in the $C$ -- string. This place we denote by
$[\delta_1, \delta_2]$, $\delta_2>\delta_1$ in \eqref{ups1} with
two $\go$'s and by $[\delta]$ in \eqref{ups0} with a single $\go$.

To simplify the subsequent star product calculation we will use
\be\label{L0sh}
\Lambda_0\to\theta^{\al}\int_{0}^{1}\dr\tau\,\tau z_{\al}e^{i\tau
z(y+p)}
\ee
in place of $\Lambda_0$, \eqref{L0} assuming that \eqref{L0sh}
acts on the corresponding $C$ in accordance with \eqref{conv1} and
similarly
\be\label{gosh}
\go(y)\to e^{ity}
\ee
in place of $\go(y)$, \eqref{conv2}. The vertices are not
difficult to calculate at a given perturbation order using
\eqref{stnd} for solving for $W$ in \eqref{Weq}. In so doing it is
convenient to set $\phi=0$ in \eqref{Wgen}. This way one finds
\be\label{Wst}
W^{(n)}=-\Delta_0\left(\{W^{(n-1)}, \Lambda_0\}_*\right)\,.
\ee
Note that contribution $\dr_x\Lambda_0$ trivially vanishes and
$W^{(n)}$ equals zero at $z=0$. It makes it convenient finding the
$z$ -- independent vertices in \eqref{verww} by setting $z=0$ in
each of the two terms. Since $\dr_x W^{(n)}_{z=0}=0$ one finds
\be
\Upsilon(\go,\go, \underbrace{C,\dots,
C}_n)=-\left(\sum_{i+j=n}W^{(i)}*W^{(j)}\right)\Big|_{z=0}\,.\label{verwst}\\
\ee
As an illustration, consider the lower order examples. Using
\eqref{W0} and \eqref{Wst} we have at the first order
\begin{align}
&W^{(1)}_{\go C}=-t^{\al}\int_{[0,1]^2}\dr\tau\dr\rho\, e^{i\rho\,
ty+i(1-\rho)pt}\hf (1-\tau)z_{\al}e^{i\tau
z(y+p+t)}\,,\label{W1}\\
&W^{(1)}_{C\go}=-t^{\al}\int_{[0,1]^2}\dr\tau\dr\rho\, e^{i\rho\,
ty+i(1-\rho)pt}\hf (1-\tau)z_{\al}e^{i\tau z(y+p-t)}\,.\label{W11}
\end{align}
Recall again that the above expressions generate $W^{(1)}$ by
acting on $\go\bar\star C$ and $C\bar\star\go$ in accordance with
\eqref{conv1}, \eqref{conv2}. Substituting it to \eqref{verwst} we
obtain up to terms  quadratic in $C$
\be
\Upsilon(\go, \go,
C)=\left(\Phi^{[1,2]}(\go\bar{\star}\go\bar{\star}C)+
\Phi^{[1,3]}(\go\bar{\star}C\bar{\star}\go)+
\Phi^{[2,3]}(C\bar{\star}\go\bar{\star}\go)\right)\Big|_{y_1'=y_2'=y_3'=0}\,,
\ee
where
\begin{align}
&\Phi^{[1,2]}=-\left(e^{i\, t_{1}\,y}*W^{(1)}_{\go
C}\right)\Big|_{z=0}\\
&\Phi^{[1,3]}=-\left(e^{i\, t_1\,y}*W^{(1)}_{C\go }+W^{(1)}_{\go
C}*e^{i\,
t_{2}\,y}\right)\Big|_{z=0}\\
&\Phi^{[2,3]}=-\left(W^{(1)}_{C\go}*e^{i\,
t_{2}\,y}\right)\Big|_{z=0}
\end{align}
or explicitly,
\begin{align}
&\Phi^{[1,2]}=t_2 t_1\int_{[0,1]^2}\dr\tau\dr\rho\,
(1-\tau)e^{i(1-\tau)(t_1+\rho\, t_2)y-i(\tau
t_1+(1-\rho)\,t_2+\tau\rho\, t_2)p
+i(\tau+(1-\tau)\rho)t_2 t_1}\,,\label{wwC}\\
&\Phi^{[1,3]}=t_1 t_2\int_{[0,1]^2}\dr\tau\dr\rho\,
(1-\tau)e^{i(1-\tau)(t_2+\rho\, t_1)y-i(\tau
t_2+(1-\rho)\,t_1+\tau\rho\,t_1)p
+i((1-\tau)\rho-\tau)t_2 t_1}+\label{wCw}\\
&+t_2 t_1\int_{[0,1]^2}\dr\tau\dr\rho\,
(1-\tau)e^{i(1-\tau)(t_1+\rho\, t_2)y-i(\tau
t_1+(1-\rho)\,t_2+\tau\rho\,t_2)p
+i((1-\tau)\rho-\tau)t_2 t_1}\,,\nn\\
&\Phi^{[2,3]}=t_1 t_2\int_{[0,1]^2}\dr\tau\dr\rho\,
(1-\tau)e^{i(1-\tau)(t_2+\rho\, t_1)y-i(\tau
t_2+(1-\rho)\,t_1+\tau\rho\,t_1)p +i(\tau+(1-\tau)\rho)t_2
t_1}\,.\label{Cww}
\end{align}
Eqs. \eqref{wwC}-\eqref{Cww} naturally reproduce a generalization
of the central on-mass-shell theorem for an arbitrary HS
background obtained in \cite{Didenko:2018fgx}. Their exponential
parts carry neither $pp$ nor $yp$ contractions. Therefore, these
vertices are ultra-local.

The lowest order form of \eqref{eq2} corresponding to the
vanishing of its right hand side comes from \eqref{Cver} as
follows. Plugging \eqref{W0} into \eqref{Cver} one gets
\begin{align}
\dr_x
C*\gga=\dr_z(\go*\Lambda_0+\Lambda_0*\go)=-\go*C*\gga+C*\gga*\go\,,
\end{align}
where we used \eqref{Lambdaeq} and the Leibniz rule, which can be
applied since $\go$ is $z$ -- independent. Noting then that
\be
\gga*\go=\pi(\go)*\gga\,,
\ee
where $\pi$ was defined in \eqref{pi}, we have at this order
\be
\dr_x C+\go*C-C*\pi(\go)=0\,.
\ee
To arrive further at the leading order one substitutes \eqref{W1},
\eqref{W11} into \eqref{Ceq} and use \eqref{awe} to find
\be
\Upsilon(\go,C,C)=\left(\Phi^{[1]}(\go\bar{\star}C\bar{\star}C)+
\Phi^{[2]}(C\bar{\star}\go\bar{\star}C)+
\Phi^{[3]}(C\bar{\star}C\bar{\star}\go)\right)\Big|_{y_1'=y_2'=y_3'=0}
\ee
with
\begin{align}
&\Phi^{[1]}*\gga=\dr_z\left(W^{(1)}_{\go C}*\Lambda_0\right)\,,\\
&\Phi^{[2]}*\gga=\dr_z\left(W^{(1)}_{C\go}*\Lambda_0+\Lambda_0*W^{(1)}_{\go
C}\right)\,,\\
&\Phi^{[3]}*\gga=\dr_z\left(\Lambda_0*W^{(1)}_{C\go}\right)\,.
\end{align}
Explicit expressions can be found using \eqref{appLf},
\eqref{appfL} and \eqref{appid} with the final result being
\begin{align}
&\Phi^{[1]}=ty\int_{[0,1]^2}\dr\rho\,\dr\gs\,\gs e^{i(\gs\rho p_2+
(1-\gs\rho)p_1)t+i(\gs p_2+(1-\gs)p_1+(1-\gs+\gs\rho)t)y}\,,\\
&\Phi^{[2]}=ty\int_{[0,1]^2}\dr\rho\,\dr\gs\,\gs e^{i(\gs\rho p_2+
(1-\gs\rho)p_1)t+i(\gs p_2+(1-\gs)p_1-(1-\gs-\gs\rho)t)y}+\nn\\
&\qquad+yt\int_{[0,1]^2}\dr\rho\,\dr\gs\,\gs e^{i(\gs\rho p_1+
(1-\gs\rho)p_2)t+i(\gs p_1+(1-\gs)p_2+(1-\gs-\gs\rho)t)y}\,,\\
&\Phi^{[3]}=yt\int_{[0,1]^2}\dr\rho\,\dr\gs\,\gs e^{i(\gs\rho p_1+
(1-\gs\rho)p_2)t+i(\gs p_1+(1-\gs)p_2-(1-\gs+\gs\rho)t)y}\,.
\end{align}
These vertices coincide with those found earlier in
\cite{Didenko:2018fgx} modulo the change of the integration over
$\gs$ and $\rho$ to the integration over a two-dimensional
simplex. Note there are no $p_1p_2$ contractions within the
exponentials meaning that the result is spin-local. Similarly one
can go on to higher orders in this fashion.

\subsection{Locality}\label{loc}
While it is not difficult to come up with expressions for vertices
$\Upsilon$ reproduced via the standard homotopy \eqref{stnd}, some
notable features are accessible without the detailed calculation.
It is easy to see that any order vertex from \eqref{eq1} is
ultra-local. This can be reached in two steps. First, we note that
\eqref{stnd} applied for solving $W^{(n)}$ leads to the following
schematic result
\be\label{Wn}
W^{(n)}\sim\sum_{\delta}\int D\rho\, t^{\al_1}\dots
t^{\al_n}e^{i\rho_y t^{\al}y_{\al}+i t^{\al}P_{t\al}}\hf
\int_{0}^{1}\dr\tau\ff{1-\tau}{\tau}\tau z_{\al_1}\dots\tau
z_{\al_n}e^{i\tau z_{\al}(y-P_z)^{\al}}\,,
\ee
where we recall that \eqref{Wn} acts on the string of $n$ $C$'s
with one impurity $\go$ at certain place $\delta$. The sum is
taken over all possible positions of $\go$. Here $\int D\rho$
denotes all repeated integrations that show up in the process of
applying $\eqref{stnd}$ except for the single one over $\tau$.
$P_z$ and $P_t$ are the linear combinations with $\rho$ --
dependent coefficients of different $p$'s
\begin{equation}
P_t=a_1(\rho)p_1+a_2(\rho)p_2+\ldots +a_n(\rho)p_n,\;\;
P_z=b_1(\rho)p_1+b_2(\rho)p_2+\ldots +b_n(\rho)p_n.
\end{equation}
While $P_t$ and $P_z$ depend on $\delta$, we deliberately ignore
any particular order of $\omega$ and $C$ since it is not important
for our conclusion of locality. Note that \eqref{Wn} is free from
nonlocalities as the only contraction that appears after $\hf$ --
computation \eqref{hf} is between $\omega$ and $C$ but never
between two $C$'s. Moreover, since there are no contractions
between $p$'s and $y$ in the left hand exponential of \eqref{Wn},
the whole expression \eqref{Wn} is ultra-local having no
occurrence of $y$ in $C$'s. Now, the vertex from \eqref{eq1} is
given by \eqref{verwst} which contains star products of $W$'s from
different orders. These star products remain ultra-local as
follows from \eqref{00}. Indeed, taking some $W^{(i)}$ from the
$i^{th}$ -- order
\be
W^{(i)}\sim\int D\rho^{(i)}\, \left(t\cdot\ff{\p}{\p
P_{z}^{(i)}}\right)^i \left(e^{i\rho_y^{(i)} t^{\al}y_{\al}+i
t^{\al}P_{t\al}^{(i)}}\hf
\int_{0}^{1}\dr\tau\ff{1-\tau}{\tau}e^{i\tau
z_{\al}(y-P_z^{(i)})^{\al}}\right)\,,
\ee
one finds
\begin{align}
&W^{(i)}*W^{(j)}\sim\nn\\
&\sim\int D\rho^{(i)}\,D\rho^{(j)}\,
\ff{\dr\gs}{\gs(1-\gs)}\left(t_1\cdot\ff{\p}{\p
P_{z}^{(i)}}\right)^i
\left(t_2\cdot\ff{\p}{\p P_{z}^{(j)}}\right)^j\times\\
&\times e^{i(\rho_y^{(i)}+\rho_y^{(j)}) t^{\al}y_{\al}+i
t^{\al}(P_{t}^{(i)}+P_{t}^{(j)})_{\al}}\hf
\int_{0}^{1}\dr\tau\ff{1-\tau}{\tau}e^{i\tau
z_{\al}(y-P_z^{(i,j)})^{\al}}\,,\nn
\end{align}
where
\be
P_{z}^{(i,j)}=\gs(P_z^{(i)}-\rho^{(j)}_yt)+(1-\gs)(P_z^{(j)}+\rho^{(i)}_yt)\,.
\ee
The above star product brings no $y\cdot p$ terms into the
exponential leaving all $C$'s $y$ -- independent in each
contribution of \eqref{verwst}. Therefore the final result
\eqref{verwst} is ultra-local. This proves ultra-locality of
\eqref{eq1} in the holomorphic sector.

Similarly, the holomorphic part of the 0-form vertices \eqref{eq2}
can be extracted from \eqref{Cver}. Taking \eqref{Wn} and
substituting it into \eqref{Cver} up to an irrelevant order of
$\go$ in the string we obtain
\be
\dr_z (W^{(n-1)}*\Lambda_0)=\Phi_L(y)*\gga\,,
\ee
where
\begin{align}\label{PhiC}
&\Phi_L(y)=\int D\rho \int_0^1 \dr\sigma \,
\frac{1-\sigma}{\sigma}\sigma^{n-1} (y^\beta t_\beta)^{n-1}\,
\exp\Big\{iy^\alpha(-\sigma P_z^{(n-1)}+(1-\sigma) \rho_y^{(n-1)} t+(1-\sigma)p_n)^\alpha+\nn\\
&+it^\alpha P_{t\alpha}^{(n-1)}+i(\sigma
P_z^{(n-1)}-\rho_y^{(n-1)}(1-\sigma)p_n)^\alpha  t_\alpha\Big\}\,.
\end{align}
Analogously with $\dr_z (\Lambda_0*W^{(n-1)})=\Phi_R(y)*\gga$.
Note that \eqref{PhiC} contains $y\cdot p$ contraction within the
exponential. Therefore, the corresponding vertex is not
ultra-local. However, since it has no $p\cdot p$ contractions it
is still spin local. This makes the holomorphic part of vertices
in \eqref{eq2} spin-local.

\subsection{Shift symmetry}\label{shift}
Now we want to specify the dependence of $P_t$ and $P_z$ on $\rho$
variables. This dependence has the following remarkable properties
\begin{equation}\label{Pt}
P_t(p_1+a,p_2+a,\ldots,p_n+a)=P_t(p_1,p_2,\ldots,p_n)+(1-\rho_y)a,
\end{equation}
\begin{equation}\label{Pz}
P_z(p_1+a,p_2+a,\ldots,p_n+a,t)=P_z(p_1,p_2,\ldots, p_n,t)-a.
\end{equation}
Here $a$ is an arbitrary spinor. One can prove that the following
properties indeed take place by induction.

It is easy to see that the base of induction, namely $W^{(1)}$,
respects such property. For example,
\begin{equation}\label{W1wC}
W^{(1)}_{\go C}=\int_0^1 d\rho\, e^{i\rho\, y^\alpha
t_\alpha+i(1-\rho)p_\alpha t^\alpha} t^\beta \hf \int_0^1 d\tau\,
\frac{1-\tau}{\tau} \tau z_\beta e^{i\tau
z_\alpha(y+p+t)^\alpha}\,.
\end{equation}
Similarly with $W^{(1)}_{C\go}$.  We already know that at any
order of perturbation theory $W^{(n)}$ acquires the form
\eqref{Wn}. Assuming that properties \eqref{Pt},\eqref{Pz} are
satisfied for the $n$-th order we are going to show that they are
satisfied for the $(n+1)$ order. Straightforward computation
yields
\begin{multline}
\Delta_0\big(W^{(n)} \ast\Lambda_0\big)=\\
=\int {D}\rho\, \rho_y \int_0^1 d\sigma \, \frac{1-\sigma}{\sigma}
\sigma^{n}\int_0^1 d\zeta\,
e^{i(1-\zeta)\rho_y y^\alpha  t_\alpha+it^\alpha
(P_{t\alpha}-\rho_y \zeta (
\sigma P_z-(1-\sigma)p_{n+1}))_\alpha}{t^{\beta_1}\ldots t^{\beta_{n+1}}}\hf\\
\\ \hf\int_0^1 d\tau\, \frac{1-\tau}{\tau}\, {\tau z_{\beta_1}
\ldots \tau z_{\beta_{n+1}}}\, e^{i\tau z_\alpha(y-\sigma
P_z+(1-\sigma)\rho_y t+(1-\sigma)p_{n+1})^\alpha}\,.
\end{multline}
From the above expressions one immediately sees that provided
properties \eqref{Pt} and \eqref{Pz} are satisfied for order $n$
they are satisfied for order $n+1$.

Such a symmetry of the master field $W$ induces the corresponding
symmetric properties on the vertices. To compute the zero-form
vertices one uses \eqref{Cver}. Straightforward computation yields
\begin{multline}
\dr_z\big(W^{(n)}\ast \Lambda\big)=\ff12\theta^2 \int D\rho
\int_0^1 d\sigma\, \frac{1-\sigma}{\sigma} \sigma^n (z_\beta
t^\beta)^n\,
\exp\Big\{iz_\alpha(y-\sigma P_z+(1-\sigma)\rho_y t+(1-\sigma) p_{n+1})^\alpha+\\
+it^\alpha P_{t\alpha}+i(\sigma P_z-(1-\sigma)p_{n+1})^\alpha
\rho_y t_\alpha\Big\}.
\end{multline}
To extract vertices for the zero-form sector one has to rewrite
this expression in the form
\begin{multline}\label{shver}
\dr_z (W^{(n)}\ast \Lambda_0)=\int D\rho \int_0^1 d\sigma \,
\frac{1-\sigma}{\sigma}\sigma^{n} (y^\beta t_\beta)^n\,
\exp\Big\{iy^\alpha(-\sigma P_z+(1-\sigma) \rho_y t+(1-\sigma)p_{n+1})^\alpha+\\
+it^\alpha P_{t\alpha}+i(\sigma P_z-(1-\sigma)p_{n+1})^\alpha
\rho_y t_\alpha\Big\} \ast \gamma.
\end{multline}
From the last expression one derives the following property of the
vertex
\begin{equation}\label{sh1}
\boxed{\Phi^{[\gd]}(y; t, p_i+a)=e^{i(t+y)^\alpha
a_\alpha}\Phi^{[\gd]}(y; t, p_i)}
\end{equation}

Now consider vertices in the one-form sector. Contributions to
these vertices come from various products $W^{(n)}\ast
W^{(n^\prime)}$. Straightforward computation yields
\begin{multline}\label{shver2}
(W^{(n)}\ast W^{(n^\prime)})|_{z=0}=\Bigg(\int D\rho\, e^{i\rho_y
y^\alpha t_\alpha+it^\alpha P_{t\alpha}}t^{\beta_1}\ldots
t^{\beta_n}\hf \int_0^1 d\tau\,
\frac{1-\tau}{\tau}\tau z_{\beta_1}\ldots z_{\beta_n}\, e^{i\tau z_\alpha (y-P_z)^\alpha}\ast \\
\ast \int D\rho^\prime\, e^{i\rho_y^\prime y^\alpha t^\prime_\alpha+
it^{\prime \alpha}P^\prime_{t\alpha}}t^{\prime \sigma_1}\ldots
t^{\prime \sigma_{n\prime}}\hf \int_0^1 d\tau^\prime \frac{1-\tau^\prime}{\tau^\prime}
\tau^\prime z_{\sigma_1}\ldots z_{\sigma_{n^\prime}}\,
e^{i\tau^\prime z_\alpha (y-P_z^\prime)^\alpha}\Bigg)\Big|_{z=0}=\\
=\int D\rho\int D\rho^\prime \, \int_0^1 d\sigma\frac{1}{\sigma(1-\sigma)}\int_0^1 d\tau\,
\frac{1-\tau}{\tau}\, (\tau\sigma \rho_y^\prime)^n((1-\sigma)\rho_y\tau)^{n^\prime}
(t_\alpha t^{\prime\alpha})^{n+n^\prime}\times\\
\times \exp\Big\{i(1-\tau)y^\alpha(\rho_y t+\rho_y^\prime t^\prime)_\alpha-i\rho_y
\rho_y^\prime t_\alpha t^{\prime \alpha}+it^\alpha P_{t\alpha}+it^{\prime \alpha}P^\prime_{t\alpha}+\\
+i\tau(\sigma P_z+\sigma \rho_y^\prime
t^\prime+(1-\sigma)P_z^\prime-(1-\sigma)\rho_y t)^\alpha(\rho_y
t+\rho_y^\prime t^\prime)_\alpha\Big\}
\end{multline}
Using \eqref{Pt} and \eqref{Pz} one can derive the following
property of the vertices for one-forms
\begin{equation}\label{sh2}
\boxed{\Phi^{[\gd_1, \gd_2]}(y-a; t_1, t_2,
p_i+a)=e^{i(t_1+t_2)^\alpha a_\alpha}\Phi^{[\gd_1, \gd_2]}(y; t_1,
t_2, p_i)\,.}
\end{equation}
Taking $a=\nu(t+y)$ in \eqref{sh1} and $a=\chi(t_1+t_2)$ in
\eqref{sh2}, where $\nu$ and $\chi$ are arbitrary numbers one
notes that
\begin{align}
&\Phi^{[\gd]}\left(y; t, p_i+\nu (t+y)\right)=\Phi^{[\gd]}(y; t, p_i)\,,\\
&\Phi^{[\gd_1\,\gd_2]}(y-\chi(t_1+t_2); t_1, t_2, p_i+\chi
(t_1+t_2))=\Phi^{[\gd_1\,\gd_2]}(y; t_1, t_2, p_i)\,.
\end{align}
The differential version of symmetries \eqref{sh1} and \eqref{sh2}
reads
\begin{align}
&\left(t+y-i\sum_j\ff{\p}{\p p_{j}}\right)\Phi^{[\delta]}(y; t,
p_i)=0\,,\\
&\left(t_1+t_2+i\ff{\p}{\p y}-i\sum_j\ff{\p}{\p
p_{j}}\right)\Phi^{[\delta_1,\, \delta_2]}(y; t_1, t_2, p_i)=0\,,
\end{align}
A few comments on the shift symmetry are in order. Its action both
in \eqref{sh1} and \eqref{sh2} is somewhat 'off-shell' in a sense
that it remains valid for the integrands of eqs. \eqref{shver},
\eqref{shver2}. Such a behavior is reminiscent of the structure
lemma from \cite{Gelfond:2018vmi} which controls parameters of the
homotopy operators used in the perturbation theory and underlies
the so called Pfaffian locality theorem. While this lemma relies
heavily on the Vasiliev star product \eqref{star} and besides that
is not about any symmetries at all, we showed that similar
shifts\footnote{The homotopy shifts of the structure lemma
\cite{Gelfond:2018vmi} include sign flips which are not present in
our case. This difference is artificial and is related to the
presence of outer Klein operator $k$ within the formalism of
\cite{Gelfond:2018vmi} which induces the aforementioned sign
alteration.} of parameters $p_i$ (accompanied by the shift $y\to
y-a$ in the sector of 1-form \eqref{sh2}) lead to the exact
symmetry of vertices based on star product \eqref{limst}. Another
comment is as shown in \cite{DK}, by postulating symmetry
\eqref{sh1} one can prove the $Z$ -- dominance conjecture from
\cite{Gelfond:2018vmi}. This implies that the observed symmetry
\eqref{sh1} and \eqref{sh2} is strongly intertwined with HS
locality.

\section{Conclusion}\label{concl}

The main result of our paper is the generating system
\eqref{dxWeq}-\eqref{Ceq} for order by order corrections of
higher-spin interactions in the holomorphic sector. The obtained
equations are of the Vasiliev type in a sense of being based on
the zero-curvature condition of a certain large algebra containing
HS algebra as a subalgebra. The original Vasiliev algebra is a
square of the HS one. It is constructed by introducing the
noncommutative variables $z$, while in our case the auxiliary
$z$'s commute as the new algebra can be viewed as a contraction of
the Vasiliev one. The obtained equations allow us to prove the all
order locality of the holomorphic HS interactions as well as to
derive a remarkable shift symmetry \eqref{sh1}, \eqref{sh2} of the
interaction vertices.

The appearance of a different algebra that underlies
\eqref{dxWeq}-\eqref{Ceq} has been identified in
\cite{Didenko:2019xzz} while studying the locality properties of
the holomorphic interaction. It was noted there that the homotopy
operators introduced for solving the Vasiliev equations which
result in the local interactions can be re-interpreted as a
one-parameter $\gb$ -- re-ordering of the original star product
followed by a contraction $\gb\to-\infty$. This procedure
effectively leads to a different large algebra in place of the
original  Vasiliev one and unambiguously prescribes the ordering
of its generating elements via star product formula \eqref{limst}.

Quite puzzling however is even though some HS vertices were
calculated using \eqref{limst} in \cite{Didenko:2019xzz}, the
naive replacement of the Vasiliev algebra with the contracted one
used in this paper at the level of the Vasiliev equations makes no
sense beyond lower orders as we explain in section \ref{sec}. The
reason is certain star products present in the Vasiliev system get
ill-defined with star product \eqref{limst} unlike those with the
original one \eqref{star}. All that suggests that on star product
\eqref{limst} there exists a consistent Vasiliev type system that
differs from the original one in some constraints. To check if it
is really so, we used the following important observation of
\cite{Didenko:2019xzz}, \cite{DeFilippi:2019jqq}. Namely, master
fields of the Vasiliev equations that exhibit (spin)-local
vertices are invariant under $\gb$ -- re-ordering \eqref{inv}.
Taking it as a definition of the proper functional class along
with a natural requirement of being closed on the operations of
the generating equations we come up with the following results
\begin{itemize}
\item The relevant functions are those given by \eqref{cl3}. They
have a natural grading with respect to degree $\dr z\equiv\theta$
being zero-, one- and two-forms. Each sector contains important
building blocks of HS dynamics. The dynamical fields appear as
$\dr_z$ -- cohomologies from ${\mathbf{C}}^{0}$, while the
interaction vertices are generated via a special 1-form
$\Lambda_0\in {\mathbf{C}}^{1}$ and the Klein 2-form
$\gamma\in{\mathbf{C}}^{2}$.

\item Star product \eqref{star} does not respect this class unlike
\eqref{limst}, which does, provided $\theta$ -- rank of the
product is less than two, \eqref{prw}, \eqref{pri}. Otherwise the
product is ill-defined unless one of the multiplier is $z$ --
independent. That explains why \eqref{limst} makes no sense on the
Vasiliev equations at higher orders yet being applicable at the
lower ones.
\end{itemize}
In constructing our generating equations the latter fact turned
out to be of a crucial importance. Unlike the standard Vasiliev
formulation, it prevents from higher order corrections to the
auxiliary connection $\Lambda$ \eqref{Lambdaeq} responsible for
the consistency. This leaves no room for the zero-form module $B$
that naturally appears in the Vasiliev case, bringing instead a
very unusual constraint \eqref{Ceq}. This constraint is seemingly
in tension with the HS interpretation of field $C$ as $z$ --
independent by definition. Remarkably, \eqref{Ceq} turned out to
be fully consistent with its $z$ -- independence thanks to the
curious projector identity \eqref{awe}, which rests on our
functional class and a very special element from the algebra
\eqref{L0}. The identity is somehow responsible for projecting
away the $z$ -- dependence. This observation being a major result
of our work plays a central role for the consistency and allows
one reproducing \eqref{eq1}-\eqref{eq2} from the system. On a side
note it would be very interesting to understand what makes
$\Lambda_0$ so special from the algebraic point of view.

Once the system is written down and is shown to be consistent, we
briefly analyze what kind of HS vertices it delivers. Solving our
equations using the standard contracting homotopy one is able to
see that all holomorphic vertices from \eqref{eq1} are
ultra-local. Following analysis from \cite{Gelfond:2019tac} this
implies the all order space-time locality. Similarly, we find the
vertices from \eqref{eq2} to be spin-local. These results prove
the locality conjecture of \cite{Gelfond:2018vmi} in the
holomorphic sector and extend the recent analysis
\cite{Didenko:2018fgx}, \cite{Didenko:2019xzz},
\cite{Didenko:2020bxd}, \cite{Gelfond:2021two} to all orders.

Another interesting observation is a shift symmetry of HS vertices
\eqref{sh1}, \eqref{sh2}. The investigation of that kind of a
symmetry was motivated by a remarkable result of
\cite{Gelfond:2018vmi}, where the so called Pfaffian locality
theorem was proven that allows one reducing the degree of
non-locality using a class of shifted homotopies. Its base is the
so called structure lemma that prescribes parametric shifts in the
perturbative homotopy operators. The shifts are designed to keep
track of the non-local $pp$ contractions within the exponential
part of the vertices. We were able to show that this observation
has an analog in the form of the exact shift symmetry of our local
vertices. Let us also note that in \cite{DK} it is shown that the
symmetry is a sufficient ingredient for the proof of the $Z$ --
dominance conjecture of \cite{Gelfond:2018vmi}. It would be
interesting to see its $CFT$ dual realization.

\section*{Acknowledgments}
I am grateful to M.A. Vasiliev for valuable comments on the draft
of the paper and to Sasha Smirnov for a very fruitful discussion.
My particular thanks go to Anatoly Korybut for his prolific
collaboration on many related aspects of this work and enjoyable
friendly atmosphere. I am indebted to Nursultan Dosmanbetov for
drawing my attention to missing terms in \eqref{wwC}-\eqref{Cww}
and a typo in \eqref{FR} fixed in the present version. I would
also like to thank the anonymous Referee for a useful remark. This
research was supported by the Russian Science Foundation grant
18-12-00507.

\newcounter{appendix}
\setcounter{appendix}{1}
\renewcommand{\theequation}{\Alph{appendix}.\arabic{equation}}
\addtocounter{section}{1} \setcounter{equation}{0}
 \renewcommand{\thesection}{\Alph{appendix}.}

\addcontentsline{toc}{section}{\,\,\,\,\,\,\,Appendix A. Deriving
functional class}

\section*{Appendix A. Deriving functional class}\label{AppA}
Let us prove that \eqref{class} satisfies the class closure
requirements
\begin{align}
&{\mathbf{C}}^r=\{\phi(z,y; \theta):\quad
\mathcal{O}_{\gb}(\phi)=\phi\,,\quad\forall\gb\}\,,\label{cla1}\\
&{\mathbf{C}}^{r_1}*{\mathbf{C}}^{r_2}\to{\mathbf{C}}^{r_1+r_2}\,,\quad r_1+r_2<2\label{}\\
&\dz {\mathbf{C}}^{r}\to{\mathbf{C}}^{r+1}\,,\label{}
\end{align}
where $\mathcal{O}_{\gb}$ is given by \eqref{Oresc} and $r$ counts
rank of $\theta$ -- degree. Consider condition \eqref{cla1}. The
invariance it places amounts to an integral equation
\be
\int\ff{\dr u\dr v}{(2\pi)^2} \phi((1-\gb)z+v,y+\gb u;
(1-\gb)\theta)\exp (iuv)=\phi(z,y; \theta)\,.
\ee
Let us start with zero-forms $r=0$. To solve it we propose the
following ansatz
\be\label{ans}
\phi=\int_{0}^{1}\dr\tau \rho(\tau)\int\ff{\dr u^2\dr
v^{2}}{(2\pi)^2}f(\tau z+v, (1-\tau)(y+u))e^{i\tau zy+i uv}\,,
\ee
where $\rho(\tau)$ is unknown function. Applying
$\mathcal{O}_{\gb}$  to \eqref{ans} and after some algebra one
finds
\be
\mathcal{O}_{\gb}(\phi)=\int_{0}^{1}\dr\tau
\ff{\rho(\tau)}{(1-\gb\tau)^2}\int\ff{\dr u^2\dr
v^{2}}{(2\pi)^2}f\left(\ff{\tau(1-\gb)z}{1-\gb\tau}+v,
\ff{1-\tau}{1-\gb\tau}(y+u)\right)e^{i\tau zy+i uv}\,.
\ee
Making use of the integration variable change
\be
\tau\to \ff{\tau}{1-\gb+\gb\tau}\in[0,1]
\ee
one arrives at the same expression as in \eqref{ans} provided the
following functional equation on $\rho$ satisfied
\be\label{rho}
\rho\left(\ff{\tau}{1-\gb+\gb\tau}\right)=(1-\gb)\rho(\tau)\,,
\ee
which solution up to an arbitrary factor is
\be
\rho(\tau)=\ff{1-\tau}{\tau}\,.
\ee
It is easy to see that with \eqref{rho} one has
${\mathbf{C}}^0*{\mathbf{C}}^0\to{\mathbf{C}}^0$. A convenient way
seeing this is by using generating functions \eqref{t0} and
representation \eqref{hf}. Taking
\be\label{phi}
\phi=\int_{0}^{1}\dr\tau\, e^{iy A}\hf \ff{1-\tau}{\tau}\,e^{i\tau
z(y+B)}\,,
\ee
one ends up with
\be\label{00}
\phi_1*\phi_2=\int_{[0,1]^2}\dr\tau\dr\gs\ff{1}{\gs(1-\gs)}\left(
e^{iyA_1}\star e^{iy A_2} \right)\hf\ff{1-\tau}{\tau}\,e^{i\tau
z(y+B_{1,2})}\in {\mathbf{C}}^0\,,
\ee
where
\be
B_{1,2}=\gs(B_1+A_2)+(1-\gs)(B_2-A_1)\,.
\ee
Recall that \eqref{phi} is considered as the generating function
with respect to sources $A$ and $B$. Among different integrals it
provides we pick only those for which $\tau$ -- pole cancels out
(see discussion after \eqref{t0}-\eqref{t2}). In this case 'poles'
at $\gs=0$, $\gs=1$ and $\tau=0$ in \eqref{00} are fictious just
as well.
Applying now $\dr_z$ to \eqref{ans} we find
\be
\dr_z\phi=\theta\int_{0}^{1}\int\ff{\dr u^2\dr
v^{2}}{(2\pi)^2}\dr\tau (1-\tau)(y+u)f(\tau z+v,
(1-\tau)(y+u))e^{i\tau zy+i uv}\,,
\ee
which belongs to ${\mathbf{C}}^1$ from \eqref{class} for $r=1$.
Thus,
\be
\dr_z {\mathbf{C}}^0\to{\mathbf{C}}^1\,.
\ee
It is straightforward to see that such defined ${\mathbf{C}}^1$
enjoys
\be
\mathcal{O}_{\gb}({\mathbf{C}}^1)=\textnormal{inv}\,.
\ee
Again, using \eqref{limst} one makes sure that
\be
{\mathbf{C}}^0*{\mathbf{C}}^1\to {\mathbf{C}}^1\,,\quad
{\mathbf{C}}^1*{\mathbf{C}}^0\to {\mathbf{C}}^1\,.
\ee
Product ${\mathbf{C}}^1*{\mathbf{C}}^1$ however is generally
ill-defined as it gains $\tau$ -- pole even for the well-defined
multipliers which leads to divergency, see e.g., \eqref{inf11}.
Finally, one readily checks that $\dr_z {\mathbf{C}}^1$ results in
$r=2$ class from \eqref{class} which is also $\mathcal{O}_{\gb}$
-- invariant. Being a two-form, its only product which is not
identically zero is the one with functions from ${\mathbf{C}}^0$.
Generally, this product does not exist for $z$ -- dependent
${\mathbf{C}}^0$ either, see e.g., \eqref{inf02}.

\renewcommand{\theequation}{\Alph{appendix}.\arabic{equation}}
\addtocounter{appendix}{1} \setcounter{equation}{0}
\addtocounter{section}{1}
\addcontentsline{toc}{section}{\,\,\,\,\,\,\,Appendix B. Projector
identity}

\section*{Appendix B. Projector identity}\label{AppB}
Here we sketch the proof of \eqref{awe}. We need to calculate
$\dr_z(f*\Lambda_0)$ and $\dr_z(\Lambda_0*f)$, where $f\in
{\mathbf{C}}^0$ and $\Lambda_0$ is given in \eqref{L0}. A
convenient way of doing this is to represent $\Lambda_0$ as
\be\label{appL}
\Lambda_0\to\theta^{\al}\int_{0}^{1}\dr\tau\,\tau z_{\al}e^{i\tau
z(y+p)}\,,\qquad p=-i\p\,,
\ee
where this formula should be understood as a generating one for
\eqref{L0} with the help of translation operator $p$ that acts on
$C$. Similarly, we can take $f\in {\mathbf{C}}^0$, \eqref{class}
\be\label{appf}
f(z, y)=\int_{0}^{1}\dr\tau\ff{1-\tau}{\tau}\int \ff{\dr u\,\dr
v}{(2\pi)^2}\, \psi \Big(\tau z+v, (1-\tau)(y+u)\Big)e^{i\tau
zy+iuv}
\ee
using its Taylor representation for
\begin{align}
&\psi(z,y)\to e^{iyA+izB}\,,\qquad B=i\p_1\,,\quad A=i\p_2\,,\label{apppsi}\\
&\psi(z,y)\equiv e^{iyA+izB}\psi(0,0)\,.
\end{align}
where $\p_{1,2}$ act on the first and the second argument of
$\psi$ correspondingly. Doing $uv$ integral in \eqref{appf} with
$\psi$ from \eqref{apppsi} and using symbol $\hf$ from
\eqref{half} we have
\be\label{appgenf}
f\to \int_{0}^{1}\dr\tau\ff{1-\tau}{\tau} e^{iy A+iB A}\hf
e^{i\tau z(y+B)}\,,
\ee
where we recall that we consider regular functions $f(z,y)$ only
for which $\tau$ -- pole cancels out (see below
\eqref{t0}-\eqref{t2}). Now, it is not difficult to come up with
the following generating expressions
\begin{align}
&\Lambda_{0}*f\to\theta^{\al}\int_{[0,1]^2}\dr\tau\,\dr\gs\,\ff{\gs}{1-\gs}\,e^{iyA+iBA}\hf
\tau
z_{\al}e^{i\tau z(y+\gs (p+A)+(1-\gs)B)}\,,\label{appLf}\\
&
f*\Lambda_{0}\to\theta^{\al}\int_{[0,1]^2}\dr\tau\,\dr\gs\,\ff{\gs}{1-\gs}\,e^{iyA+iBA}\circ
\tau z_{\al}e^{i\tau z(y+\gs (p-A)+(1-\gs)B)}\,,\label{appfL}
\end{align}
which are obtained directly by using \eqref{limst} along with an
appropriate change of variable $\tau$. The last step is to check
that by applying $\dr_z$ to \eqref{appLf} and \eqref{appfL} the
result amounts to $F_{\pm}(y)*\gga$, \eqref{FR}. To this end we
observe the following identity
\be\label{appid}
\dr_z \int_{0}^{1}\dr\tau\,\phi(y)\hf \tau\, \theta^{\al}
z_{\al}e^{i\tau z(y+q)}=\phi(-q) e^{-iyq}*\gga\,,
\ee
which holds for any $\phi(y)$ and $q$. Eq. \eqref{appid} is not
hard to prove by using the two-component Schouten identity
\be
\theta^{\al}\theta^{\gb}=\ff12
\theta_{\gga}\theta^{\gga}\gep^{\al\gb}\,,
\ee
which allows one to get to the following expression upon carrying
out $\hf$ -- product
\be
\dr_z \int_{0}^{1}\dr\tau\,\phi(y)\hf \tau \theta^{\al}
z_{\al}e^{i\tau z(y+q)}=\ff12\theta_{\al}\theta^{\al}
e^{izy}\int_{0}^{1}\dr\tau\,\p_{\tau}\left(\tau^2
\phi(x)e^{-izx}\right)\,,\quad x=(1-\tau)y-\tau q\,.
\ee
The integral over $\tau$ is given by a total derivative which
reduces the integration to the value at $\tau=1$ thus proving
\eqref{appid}. It remains to plug \eqref{appLf} and \eqref{appfL}
into \eqref{appid} and fold up the generating functions
\eqref{appL} and \eqref{appgenf} back to obtain \eqref{FR}. In
doing so we observe that substitution of \eqref{appfL} into
\eqref{appid} upon changing $\gs=1-\tau$ can be rewritten in the
following form
\be
\dr_z(\Lambda_0*f)=\left(\int_{0}^{1}\dr\tau\ff{1-\tau}{\tau}
e^{i(1-\tau)(-y-p)A+i\tau z(-y-p+B)+i(1-\tau)B A}\Big|_{z=-y}\cdot
e^{-iyp}\right)*\gga\,.
\ee
Now we see that due to \eqref{appgenf} the integral above  is
\begin{align}
\int_{0}^{1}\dr\tau\ff{1-\tau}{\tau} e^{i(1-\tau)(-y-p)A+i\tau
z(-y-p+B)+i(1-\tau)B A}&=\\
\int_{0}^{1}\dr\tau\ff{1-\tau}{\tau} e^{i(-y-p) A+iB A}\hf
e^{i\tau z(-y-p+B)}&\to f(z,-y-p)\,.
\end{align}
Finally, one is left to note that with prescription $C(y)\to
e^{-iyp}$ and using $y$ -- star product \eqref{exp} the result
indeed amounts to \eqref{FR}
\be
f(-z,-y-p)\Big|_{z=y}e^{-iyp}\to (C(y)*_{y}f(-z,-y))\Big|_{z=y}\,.
\ee
The case of $\dr_z(f*\Lambda_0)$ is reached analogously.

\end{document}